# T-Quark Mass and

# Hyperfinite II1 von Neumann factor

Frank D. (Tony) Smith, Jr., Cartersville - July-August 2002, August-October 2003

---

Abstract:

**My theoretical model** ( which I call the D4-D5-E6-E7-E8 VoDou Physics Model , because it is based on the Lie algebras D4,D5,E6,E7,E8 and on Clifford algebra periodicity related to IFA =VoDou ) **meets Einstein's Criterion**:

"... a theorem which at present can not be based upon anything more than upon a faith in the simplicity, i.e., intelligibility, of nature: **there are no arbitrary constants** ... that is to say, nature is so constituted that it is possible logically to lay down such strongly determined laws that within these laws **only rationally completely determined constants occur (not constants, therefore, whose numerical value could be changed without destroying the theory)**. ...".

According to the model, geometry of the Hermitian Symmetric Spaces D5 / D4xU(1) and E6 / D5xU(1) and related Shilov Boundaries, along with combinatorial relations, allows the calculation of ratios of particle masses:

- Me-neutrino = Mmu-neutrino = Mtau-neutrino = 0 (tree-level)
- Me = 0.5110 MeV
- Md = Mu = 312.8 MeV (constituent quark mass)
- Mmu = 104.8 MeV
- Ms = 625 MeV (constituent quark mass)
- Mc = 2.09 GeV (constituent quark mass)
- Mtau = 1.88 GeV
- Mb = 5.63 GeV (constituent quark mass)
- **Mt = 130 GeV (constituent Truth Quark mass)**

and

- W+ mass = W- mass = 80.326 GeV
- Z0 mass = 91.862 GeV
- Higgs mass = 145.8 GeV
- weak force - Higgs VEV = 252.5 GeV (assumed, since ratios are calculated)

as well as ratios of force strength constants:

- Gravitational G = (Ggravity)(Mproton)^2 = 5 x 10^(-39) (assumed, since ratios are calculated)
- **electromagnetic fine structure constant = 1/137.03608**
- Gfermi = (Gweak)(Mproton)^2 = 1.02 x 10^(-5)
- color force strength = 0.6286 (at 0.245 GeV) - perturbative QCD running gives
  - color force strength = 0.167 (at 5.3 GeV)
  - color force strength = 0.121 (at 34 GeV)
  - color force strength = 0.106 (at 91 GeV)
- If Nonperturbative QCD and other things are taken into account, then the color force strength = 0.123 (at 91 GeV)

The theoretical calculated electromagnetic fine structure constant = 1/137.03608 solves **Feynman's mystery** (QED, Princeton 1985, 1988, at page 129): "... **the inverse of ... about 137**.03597 ... [the] square [of] ... the amplitude for a real electron to emit or absorb a real photon ... **has been a mystery** ever since it was discovered more than fifty years ago, and **all good theoretical physicists put this number up on their wall and worry about it**. ...".

**The Truth Quark constituent mass (tree-level) calculation of about 130 GeV had been made by February 1984**.

About 10 years later, in April 1994, Fermilab officially announced observation of the Truth Quark.

Fermilab's analysis of the events gives a T-quark mass of about 170 GeV.

**My independent analysis of the same Fermilab events gives a Truth Quark mass of about 130 GeV**, consistent with the theoretical tree-level calculation.

The local Lagrangian of the theoretical model is based on the structure of the real Cl(1,7) Clifford algebra which, through 8-fold periodicity, might be used to construct a Generalized Hyperfinite II1 von Neumann Algebra factor.

---

## Table of Contents:

- **Fermions.**
- **Kobayashi-Maskawa Parameters.**
- **Proton-Neutron Mass Difference.**
- **UCC - DCC Baryon Mass Difference.**

- **Root Vector Geometry of Fermions, SpaceTime, Gauge Bosons, and D4-D5-E6-E7-E8.**

---

# Truth Quark Experimental Results.

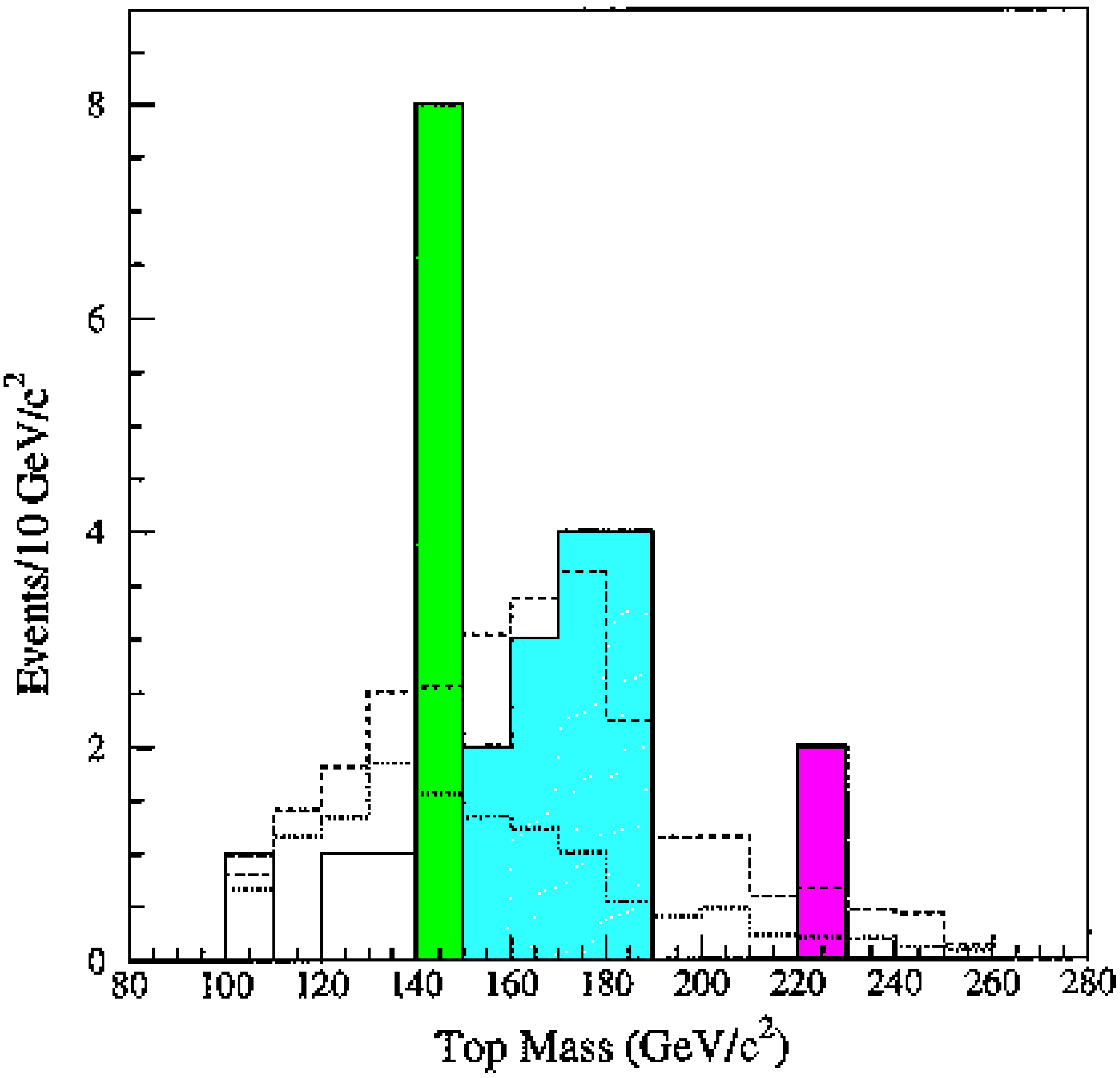

In April 1994, CDF at Fermilab (in FERMILAB-PUB-94/097-E) reported a T-quark mass of 174 (+/-10)(+13/-12) GeV. The data analyzed by CDF included a 26-event histogram for Semileptonic events with W + (3 or more) jets, without b-tags, which is Figure 65 of the report. In the histogram, the green bars are in the 140-150 GeV bin, close to the 130 GeV range that is deemed insignificant by Fermilab's analysts, but considered by me to represent the Truth quark; the cyan-blue bars represent bins in the 173 GeV range containing Semileptonic events interpreted by Fermilab as Truth Quarks; and the magenta bars represent bins in the 225 GeV range containing Semileptonic events not considered by Fermilab's analysts or by me as corresponding to either 130 GeV or 173 GeV Truth Quarks.

The peak of 8 events in the 140-150 GeV bin, shown in green, were excluded from the analysis by CDF on the grounds that (see page 140 of the report) "... the bin with masses between 140 and 150 GeV/c^2 has eight events.

We assume the mass combinations in the 140 to 150 GeV/c^2 bin represent a statistical fluctuation since their width is narrower than expected for a top signal. ...".

If the 140-150 GeV peak were only a statistical fluctuation seen by the CDF detector, one would not expect to find such a peak repeated in the data seen by the D0 detector at Fermilab. However, in March 1997, D0 (in hep-ex/9703008) reported a T-quark mass of 173.3 GeV (+/- 5.6 stat +/- 6.2 syst), based on data including a histogram similar to Figure 65 of the April 1994 CDF report which is Figure 3 of the D0 report, to which I have added colors as described above:

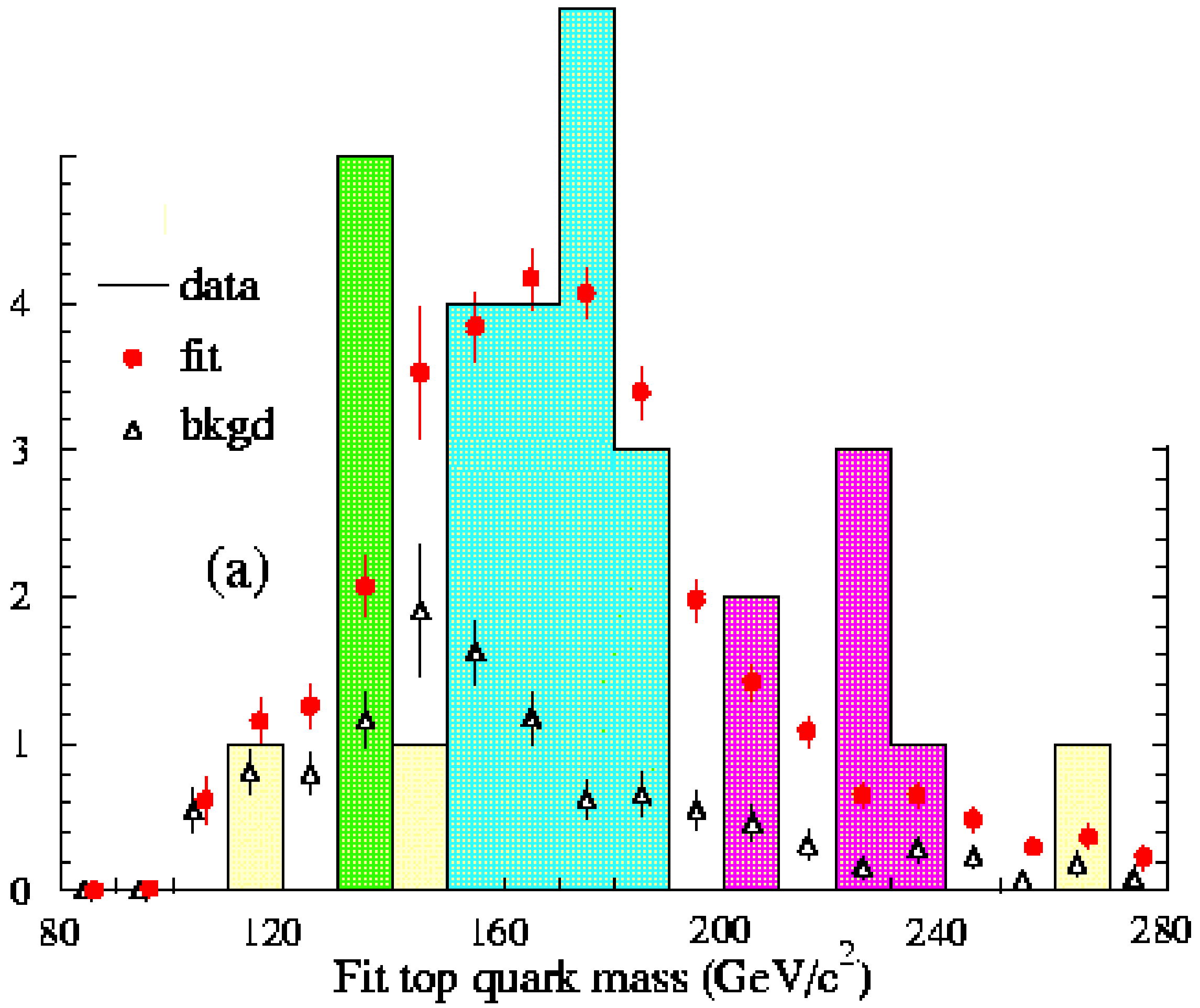

Some of the D0 histogram events, shown in cyan-blue, are are in the 150-190 GeV range and do support the CDF analysis. However, similar to the 140-150 GeV bin peak seen and thrown out by CDF, there is a peak of 5 events in the 130-140 GeV bin, shown in green, that were excluded from the analysis by D0. I did not see in the D0 report an explicit discussion of the 5-event peak in the 130-140 GeV bin.

Those 130-150 GeV peaks are from untagged semileptonic events.

---

Tagged semileptonic events may be a more reliable measure of T-quark mass, although there are fewer of them, so that statistics are not as good.

CDF (in hep-ex/9801014, dated 30 September 1997) reported a T-quark mass of 175.9 +/- 4.8(stat.) +/- 4.9(syst.) GeV based on events that were either SVX tagged, SVX double tagged, or untagged. However, CDF analysis of tagged semileptonic events (14 of them) gave a T-quark Mass of 142 GeV (+33, -14), as shown in their Figure 2, which is a plot of events/10 GeV bin vs. Reconstructed Mass in GeV:

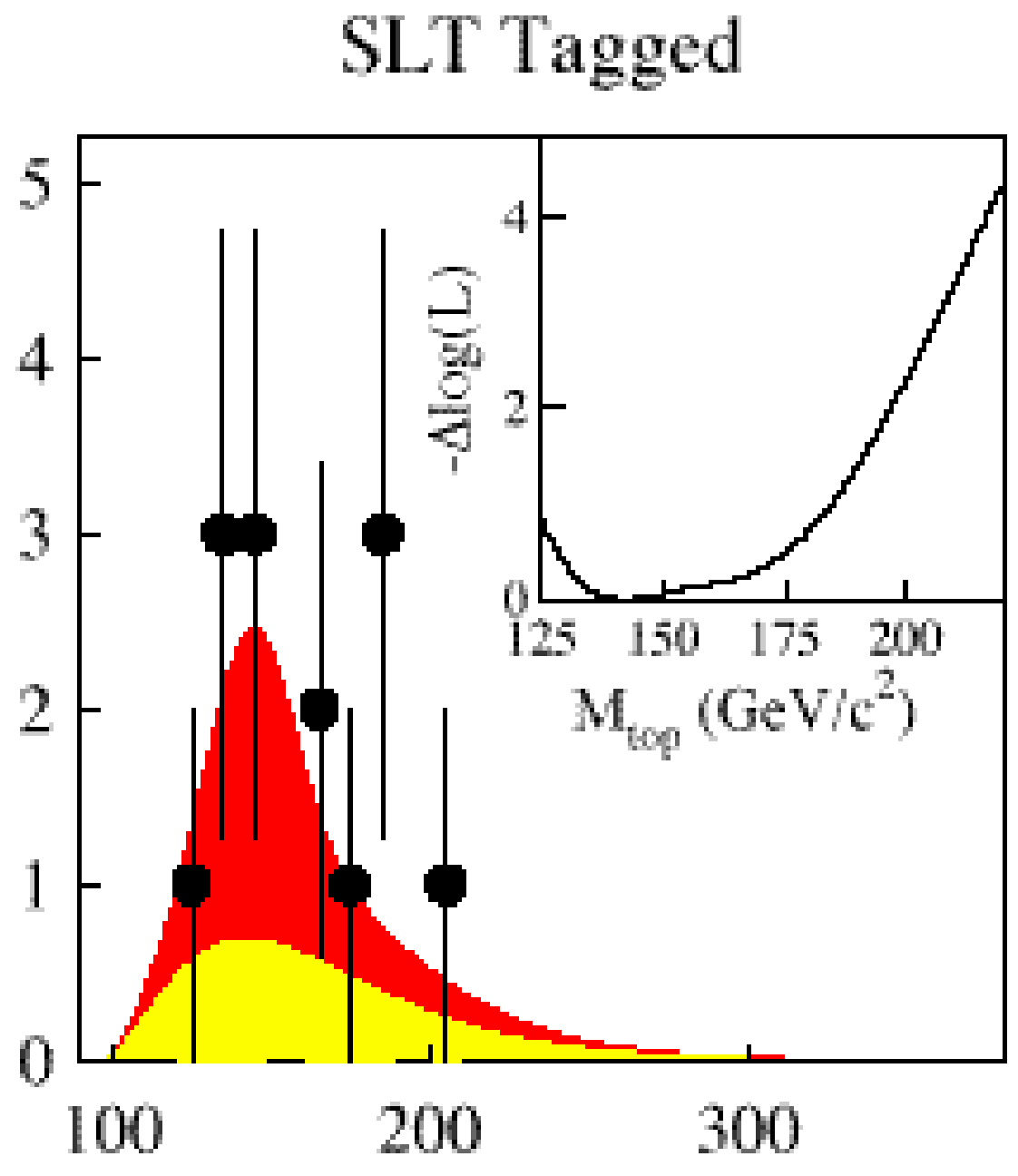

D0 (in hep-ex/9801025) also analyzed tagged semileptonic events, with the result shown in their figure 25:

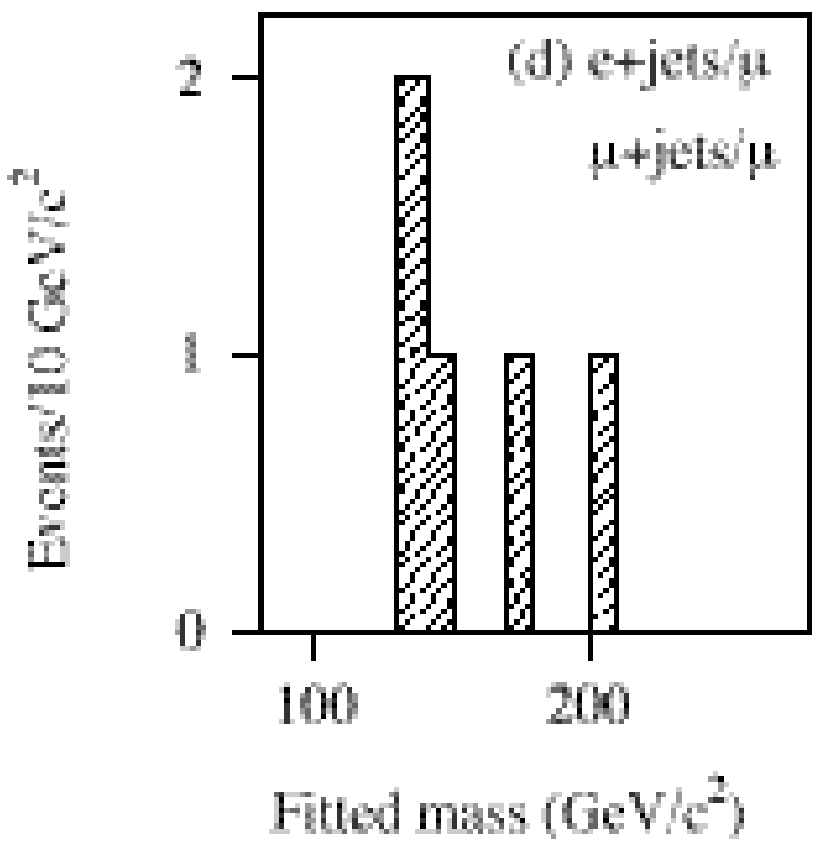

The figure shows 3 events in the 130-150 GeV range, one event in the 170-180 GeV bin, and one event in the 200-210 GeV bin. According to footnote 10 of hep-ex/9801025,

One event which would have otherwise passed the cuts, event (95653; 10822), was removed by D0 from its analysis because it was selected by the dilepton mass analysis. If this event is treated as a l + jets candidate, it has a fit Chi-squared of 0.92 and fitted Truth Quark mass of 138.7 GeV.

---

Dilepton events may be the most reliable measure of T-quark mass, although they are the least numerous type of event, so that statistics are not so good.

In October 1998 (in hep-ex/9810029) CDF analyzed 8 dilepton events and reported a T-quark mass of 167.4 +/- 10.3(stat) +/- 4.8(syst) GeV. Figure 2 of the report shows the 8 events:

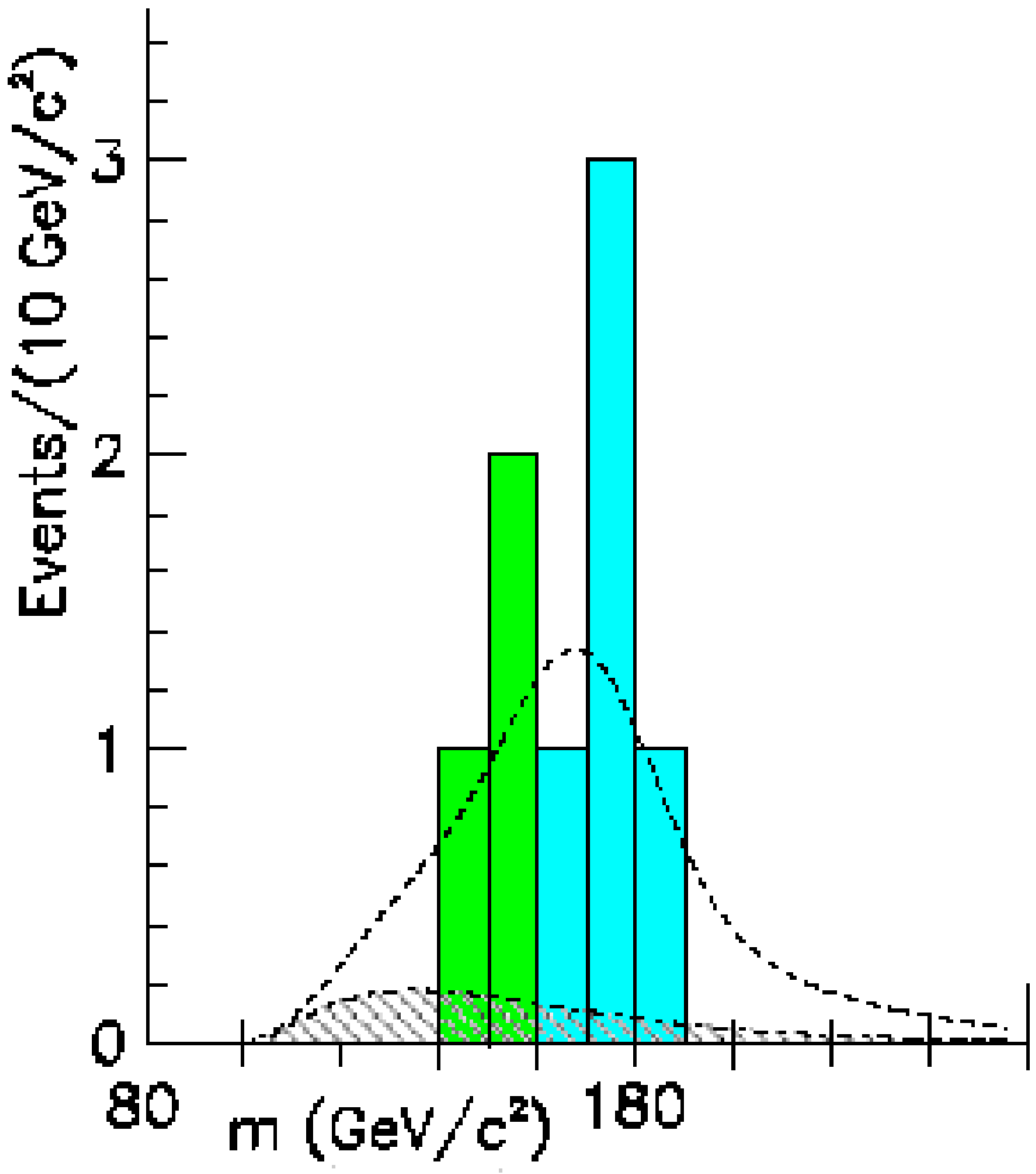

I have colored green the events with T-quark mass less than 160 GeV, and blue the events with T-quark mass greater than 160 GeV. The hep-ex/9810029 CDF report stated that it "... supersedes our previously reported result in the dilepton channel ...".

The superseded previous CDF dilepton report (hep-ex/9802017) analyzed 9 events out of a total of 11 events, which 11 events are shown on the following histogram:

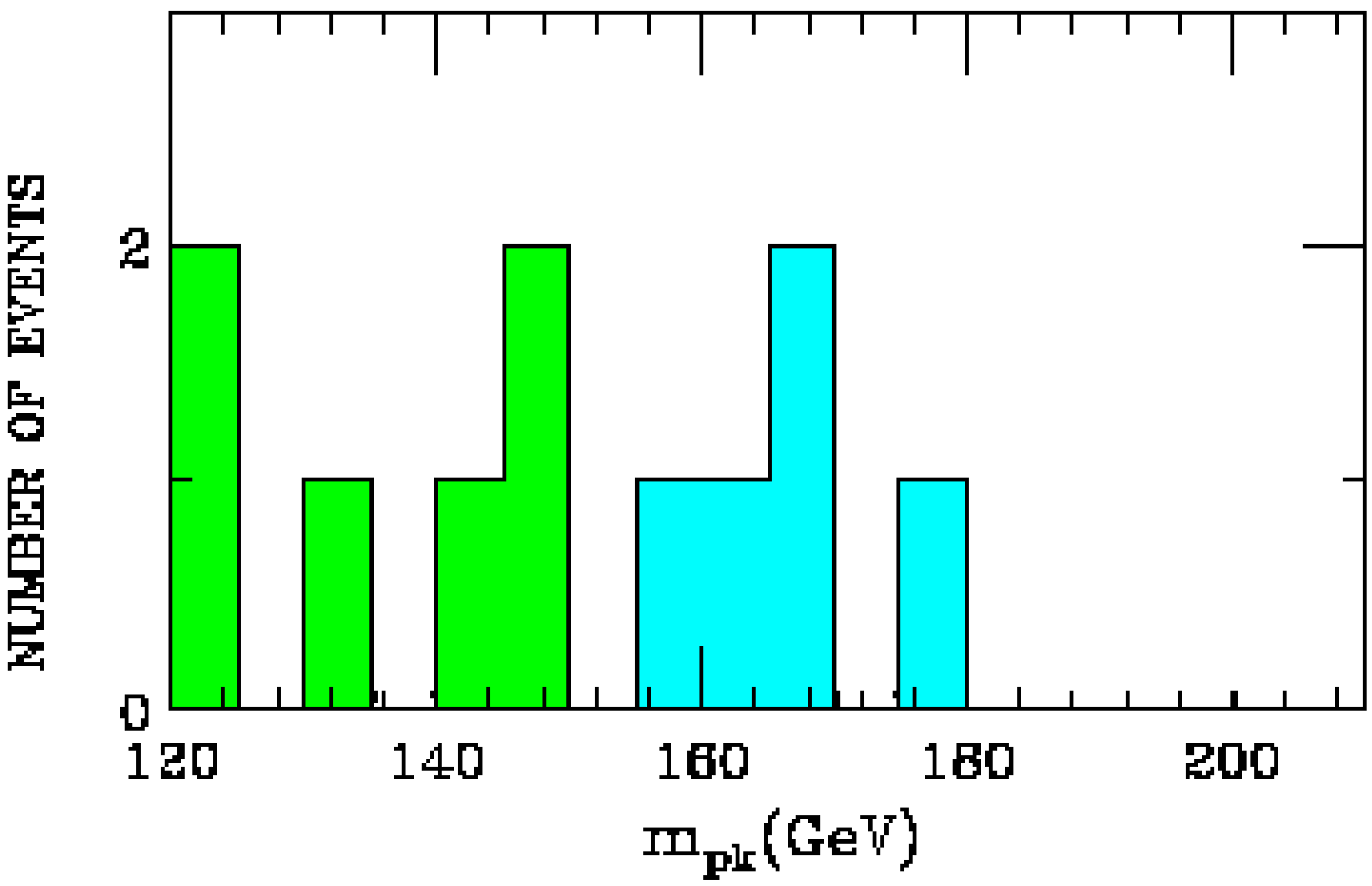

The distribution of $m_{pk}$ values determined from 11 CDF dilepton events available empirically.

I have colored green the events with T-quark mass less than 150 GeV, and blue the events with T-quark mass greater than 150 GeV.

Note first, that in the earlier 11-event histogram 5 events are shown as greater than 150 GeV, but only 4 events are shown as greater than 160 GeV, while in the 8-event revised histogram 5 events are shown as greater than 160 GeV. This indicates to me that some changes in the analysis have shifted the event mass assignments upward by about 10 GeV.

Note second, that the earlier 11-event histogram contains 3 events from 120-140 GeV that are omitted from the 8-event revised histogram.

D0 (in hep-ex/9706014 and hep-ex/9808029) has analyzed 6 dilepton events, reporting a T-quark mass of about 168.4 GeV. The 1997 UC Berkeley PhD thesis of Erich Ward Varnes which can be found on the web at http://wwwd0.fnal.gov/publications_talks/thesis/thesis.html contains details of the events and the D0 analyses. Each of the 6 events has its own characteristics. In this letter I will only discuss one of them, Run 84676 Event 12814, an electron-muon dilepton event. This figure from page 159 of the Varnes thesis, shows a T-quark mass likelihood plot calculated by the neutrino weighting algorithm.

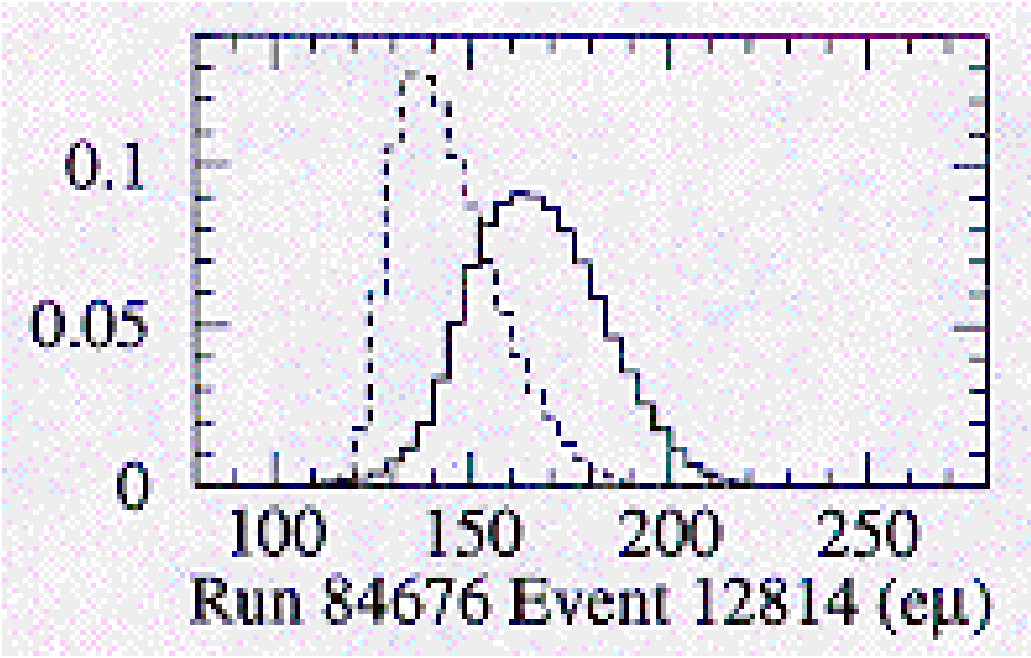

In this event there were 3 jets instead of the 2 jets you would normally expect in a Dilepton event.

The solid line is the plot if all 3 jets are included, and the dashed line is the plot if only 2 of the jets are included by excluding the third (lowest transverse energy) jet.

The 3-jet interpretation supports the 170 GeV mass favored by the Fermilab consensus, while

**the 2-jet interpretation supports a 130-140 GeV mass analysis that favors my calculated mass of about 130 Gev.**

**If the ground state of the Truth Quark is at 130 GeV, then what might the peaks at 173 GeV and 225 GeV represent? Could they be**

# Excited States from interactions among Truth Quark - Higgs - Vacua ?

Consider the Higgs mass - Truth Quark mass plane, based on Fig. 3 of Froggatt's paper hep-ph/0307138:

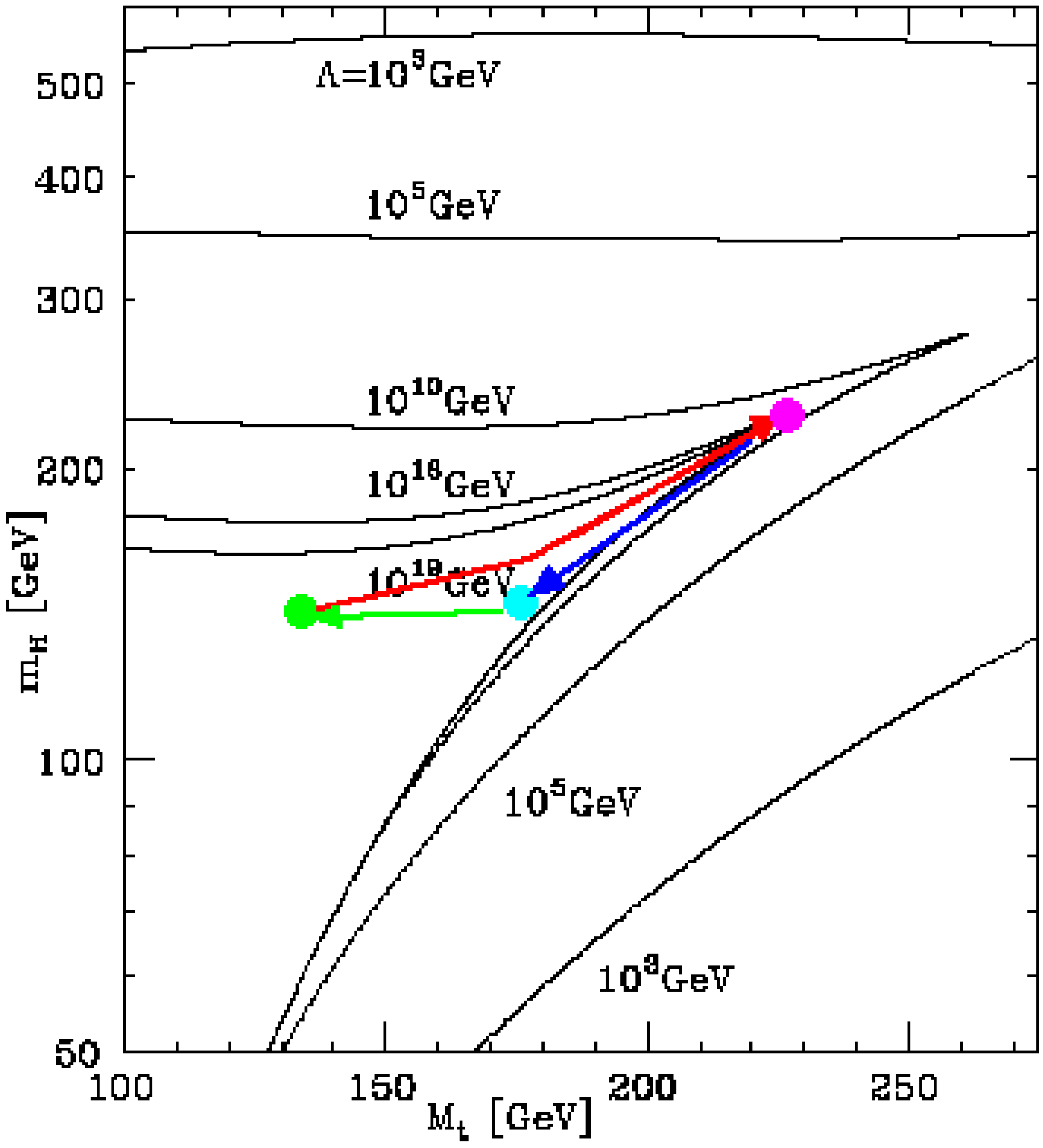

The green dot corresponds to a 130 GeV Truth Quark low-energy Standard Model one-vacuum ( < phi_vac1 > = 252 GeV ) ground state that is well within the Stability Region below the Triviality Bound and above the Vacuum Stability bound for a Standard Model with a high-energy cut-off that goes all the way to the Planck energy 10^19 GeV.

If accelerator-event collisions deposit up to 95 GeV of extra energy into a Truth Quark, it will be pumped up along the red curve within the Stability Region until it hits the Standard Model Critical Point at the magenta dot, where it will be a Standard Model Truth Quark excited state with mass-energy 225 GeV.

Since **the 225 GeV Standard Model Truth Quark excited state is at a Critical Point, which is by definition on the Vacuum Stability curve**, the one low-energy Standard Model vacuum ( < phi_vac1 > = 252 GeV ) is no longer stable, and **a new vacuum phi_vac2 forms**.

In the theoretical model, the new vacuum phi_vac2 appears at the Planck energy, where the low-energy Standard Model, Higgs, and Gravity with 4-dimensional Physical Spacetime makes a transition to a more unified structure with a Spin(1,7) gauge boson Lie algebra, fermion spinors from a Cl(1,7) Clifford algebra, an 8-dimensional Spacetime, and a corresponding high-energy vacuum with **< phi_vac2 > = 10^19 GeV = Planck energy**.

When the second vacuum phi_vac2 appears, the structure of the Standard Model is altered ( as explained by Froggatt ) so that the new Critical Point is at a 173 GeV Truth Quark mass, so the magenta dot 225 GeV excited Truth Quark state decays, moving along the blue curve along the Vacuum Stability bound to an intermediate excited state at the cyan dot at a Truth Quark mass of 173 GeV.

If the region around the Truth Quark does not have enough energy-density to maintain the second Planck energy phi_vac2 vacuum ( as is the case with present-day colliders that can do Truth Quark experiments ) then the cyan dot 173 GeV intermediate excited Truth Quark state decays along the green curve to the more stable Truth Quark low-energy Standard Model one-vacuum ground state at constituent mass of 130 GeV, the green dot.

As a result:

## The Truth Quark may provide a Window on the 8th Dimension.

There are some actual Fermilab events that seem to me to show that process in action. They are dilepton events, for which I would normally expect to see 2 jets in addition to the 2 leptons. However, some of the 6 D0 dilepton events described in the 1997 UC Berkeley PhD thesis of Erich Ward Varnes and in hep-ex/9808029 have 3 jets, and I will here discuss the kinematics of two of those events to illustrate the above-described Truth Quark - Higgs - Vacua process. The kinematics of those two events are given in Appendix B.2 of the 1997 UC Berkeley PhD thesis of Erich Ward Varnes. (Similar kinematic data are presented in D0 August 1998, hep-ex/9808029.) In the Varnes Kinematics tables, there are two numbers for each jet: one is energy after CAFIX corrections; and the second (in parentheses) is energy after post-CAFIX corrections.

The first dilepton event, Run 84395, Event 15530 ( mu mu ), as analyzed using the neutrino weighting algorithm,

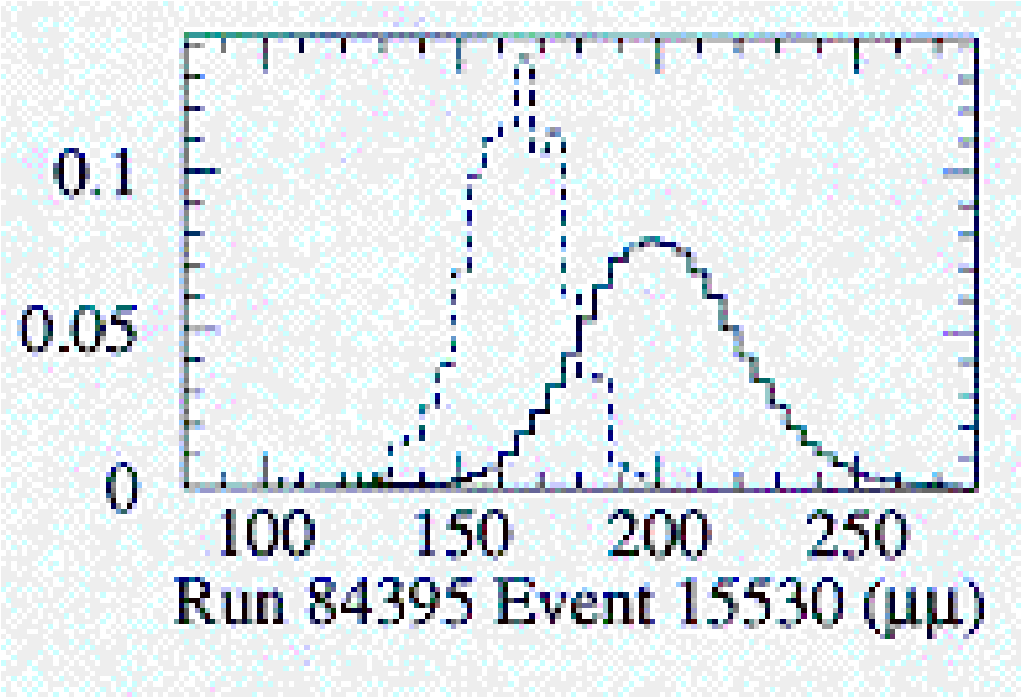

| Run 84395 Event 15530 | | | | | z vertex: 5.9 cm | | |
|---|---|---|---|---|---|---|---|
| Object | $E$ | $E_x$ | $E_y$ | $E_z$ | $E_T$ | $\eta$ | $\phi$ |
| Muon 1 | 68.6 | -63.9 | 12.7 | -21.4 | 65.1 | -0.32 | 2.94 |
| Muon 2 | 34.9 | -16.0 | 31.0 | 1.9 | 34.9 | 0.05 | 2.05 |
| $\not{E}_T$ | – | 71.2 | 53.2 | – | 88.9 | – | 0.64 |
| Jet 1 | 146.1 (153.5) | 32.1 (33.8) | -98.2 (-103.1) | -102.4 (-107.6) | 103.3 (108.5) | -0.88 | 5.03 |
| Jet 2 | 35.1 (37.2) | -8.6 (-9.1) | 21.4 (22.7) | 26.2 (27.7) | 23.1 (24.5) | 0.97 | 1.95 |
| Jet 3 | 47.1 (52.3) | -7.6 (-8.4) | -16.8 (-18.6) | 43.0 (47.8) | 18.4 (20.5) | 1.58 | 4.29 |

has, if all 3 jets are included ( the solid line in the graph ), energy around 200 GeV, corresponding to the Standard Model Critical Point Truth Quark excited state at the magenta dot. If only the 2 highest energy jets are included ( the dashed line in the graph ), it has energy around 170 GeV, corresponding to the 2-vacuum intermediate excited Truth Quark state at the cyan dot, and the energy of the third jet would correspond to the decay down the blue curve along the Vacuum Stability bound. This same event, if analyzed usingthe matrix-element weighting algorithm that, according to hep-ex/9808029, "... is an extension of the weight proposed in [R.H. Dalitz and G.R. Goldstein, Phys. Rev. D45, 1531 (1992)] ...", indicates the eventual decay into the Truth Quark low-energy Standard Model one-vacuum ground state at constituent mass of 130 GeV at the green dot.

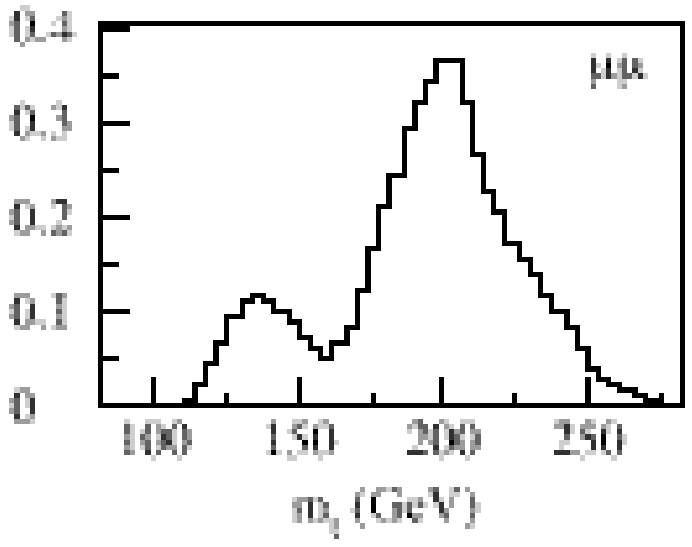

The other dilepton event that I will discuss is Run 84676, Event 12814 ( e mu ), as analyzed using the neutrino weighting algorithm:

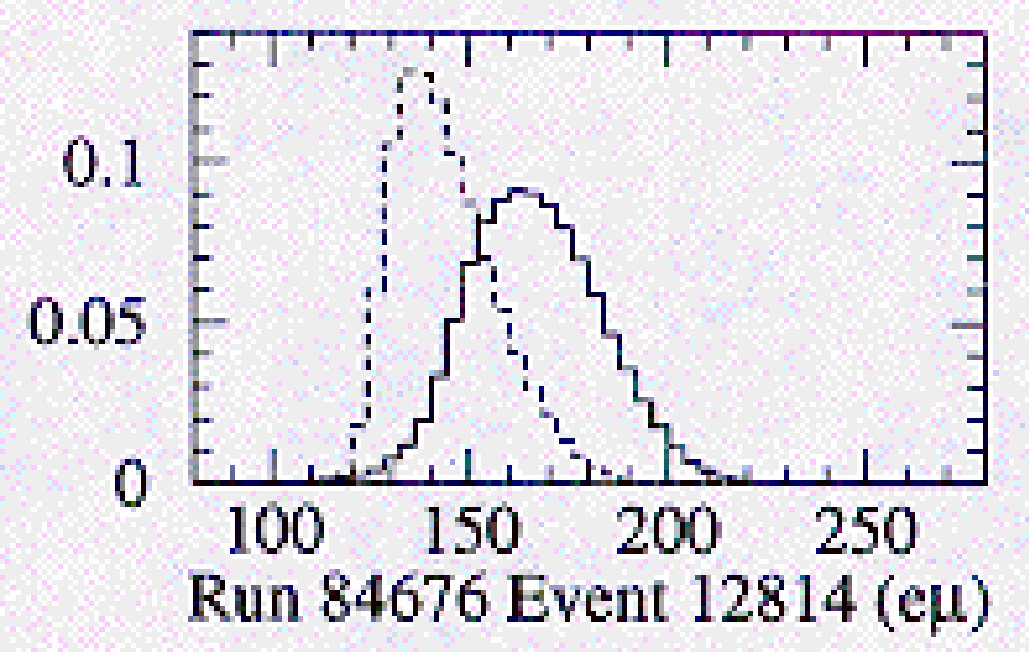

| Run 84676 Event 12814 | | | | | z vertex: -6.17 cm | | |
|---|---|---|---|---|---|---|---|
| Object | $E$ | $E_x$ | $E_y$ | $E_z$ | $E_T$ | $\eta$ | $\phi$ |
| Electron | 81.3 | -75.4 | -1.1 | -30.2 | 74.5 | -0.39 | 3.16 |
| Muon | 30.2 | -25.2 | 10.6 | -12.8 | 27.4 | -0.45 | 2.75 |
| $\not{E}_T$ | – | 62.0 | 5.2 | – | 62.3 | – | 0.08 |
| Jet 1 | 93.8 | 38.0 | -83.7 | -15.6 | 91.9 | -0.17 | 5.14 |
| | (95.9) | (38.9) | (-85.6) | (-16.0) | (94.0) | | |
| Jet 2 | 37.8 | 13.9 | 32.3 | -11.2 | 35.2 | -0.31 | 1.17 |
| | (38.8) | (14.2) | (33.1) | (-11.4) | (36.0) | | |
| Jet 3 | 31.4 | -1.6 | 28.6 | 11.6 | 28.7 | 0.39 | 1.63 |
| | (32.2) | (-1.6) | (29.3) | (11.9) | (29.4) | | |

It has, if all 3 jets are included ( the solid line in the graph ), energy around 170 GeV, corresponding to the 2-vacuum intermediate excited Truth Quark state at the cyan dot. If only the 2 highest energy jets are included ( the dashed line in the graph ), it has energy around 130 GeV, corresponding to the Truth Quark low-energy Standard Model one-vacuum ground state at constituent mass of 130 GeV at the green dot, and the energy of the third jet would correspond to the decay down the green curve.

---

## Cl(1,7) Clifford Algebra, 8-Periodicity, and a Real Hyperfinite von Neumann Algebra factor.

Complex Clifford Periodicity

Cl(2N;C) = Cl(2;C) x ...(N times tensor product)... x Cl(2;C)

Cl(2;C) = M2(C) = 2x2 complex matrices

spinor representation = 1x2 complex column spinors

Hyperfinite II1 von Neumann Algebra factor is the completion of the union of all the tensor products

Cl(2;C) x ...(N times tensor product)... x Cl(2;C)

By looking at the spinor representation, you see that "the hyperfinite II1 factor is the smallest von Neumann algebra containing the creation and annihilation operators on a fermionic Fock space of countably infinite dimension."

In other words, Complex Clifford Periodicity leads to the complex hyperfinite II1 factor which represents Dirac's electron-positron fermionic Fock space.

Now, generalize this to get a representation of ALL the particles and fields of physics.

Use Real Clifford Periodicity to construct a Real Hyperfinite II1 factor as the completion of the union of all the tensor products

Cl(1,7;R) x ...(N times tensor product)... x Cl(1,7;R)

where the Real Clifford Periodicity is

Cl(N,7N;R) = Cl(1,7;R) x ...(N times tensor product)... x Cl(1,7;R)

The components of the Real Hyperfinite II1 factor are each

Cl(1,7;R)

[ my convention is (1,7) = (-+++++++) ]

Cl(1,7) is 2^8 = 16x16 = 256-dimensional, and has graded structure

**1 8 28 56 70 56 28 8 1**

---

## D4-D5-E6 Lagrangian Structure.

**Construct the Standard Model plus Gravity Lagrangian of the theoretical model based on the structure of the Cl(1,7) Clifford Algebra. Cl(1,7) is 2^8 = 16x16 = 256-dimensional, and has graded structure**

**1 8 28 56 70 56 28 8 1**

What are the physical interpretations of its representations?

There are two mirror image half-spinors, each of the form of a real (1,7) column vector with octonionic structure.

The 1 represents:

the neutrino.

The 7 represent:

the electron;

the red, blue, and green up quarks;

the red, blue, and green down quarks.

One half-spinor represents first-genenerationfermion particles, and its mirror image represents first-generation fermion antiparticles.

Second and third generation fermions come from dimensional reduction of spacetime, so that

- first generation - octonions
- second generation - pairs of octonions

- third generation - triples of octonions

There is a (1,7)-dimensional vector representation that corresponds to an 8-dimensional high-energy spacetime with octonionic structure

that reduces at lower energies to quaternionic structures that are

- a (1,3)-dimensional physical spacetime [my convention is (1,3)=(-+++)]
- a (0,4)-dimensional internal symmetry space

There is a 28-dimensonal bivector representation that corresponds to the gauge symmetry Lie algebra Spin(1,7)

that reduces at lower energies to:

- a 16-dimensional U(2,2) = U(1)xSU(2,2) = U(1)xSpin(2,4) whose conformal Lie algebra / Lie group structure leads to gravity by a mechanism similar to the MacDowell-Mansouri mechanism;
- a 12-dimensional SU(3)xSU(2)xU(1) Standard Model symmetry group that is represented on the internal symmetry space by the structure SU(3) / SU(2)xU(1) = CP2.

There is a 1-dimensional scalar representation for the Higgs mechanism.

The Cl(1,7) Clifford Algebra structures

**1 8 28 56 70 56 28 8 1 = (8+8)x(8+8)**

fit together to form a Lagrangian in 8-dimensional SpaceTime that can be written, prior to dimensional reduction, as

the Integral over 8-dim SpaceTime of

dd P' /\ * dd P + F /\ *F + S' D S + GF + GG

where d is the 8-dim covariant derivative

P is the scalar field

F is the Spin(8) curvature

S' and S are half-spinor fermion spaces

D is the 8-dim Dirac operator

GF is the gauge-fixing term

GG is the ghost term

As shown in chapter 4 of Gockeler and Schucker,

the scalar part of the Lagrangian dd P' /\ * dd P becomes Fh8 /\ *Fh8

where Fh8 is an 8-dimensional Higgs curvature term.

After dimensional reduction to 4-dim SpaceTime, the scalar Fh8 /\ *Fh8 term becomes the Integral over 4-dim Spacetime of

(Fh44 + Fh4I + FhII) /\ *(Fh44 + Fh4I + FhII) =

= Fh44 /\ *Fh44 + Fh4I /\ *Fh4I + FhII /\ *FhII

where cross-terms are eliminated by antisymmetry

of the wedge /\ product

and 4 denotes 4-dim SpaceTime

and I denotes 4-dim Internal Symmetry Space

The Internal Symmetry Space terms should be integrated over the 4-dimensional Internal Symmetry Space, to get **3 terms.**

**The first term** is the integral over 4-dim SpaceTime of

Fh44 /\ *Fh44

Since they are both SU(2) gauge group terms, this term merges into the SU(2) weak force term that is the integral over 4-dim SpaceTime of Fw /\ *Fw (where w denotes Weak Force).

**The third term** is the integral over 4-dim SpaceTime of the integral over 4-dim Internal Symmetry Space of

FhII /\ *FhII

The third term after integration over the 4-dim Internal Symmetry Space, produces, by a process similar to the Mayer Mechanism developed by Meinhard Mayer, terms of the form

L (PP)^2 - 2 M^2 PP

where L is the Lambda term, P is the Phi scalar complex doublet term, and M is the Mu term in the wrong-sign Lamba Phi^4 theory potential term, which describes the Higgs Mechanism. The M and L are written above in the notation used by Kane and Barger and Phillips. Ni, and Ni, Lou, Lu, and Yang, use a different notation

- ( 1 / 2 ) Sigma Pn Pn + ( 1 / 4! ) Ln (PnPn)^2

so that the L that I use (following Kane and Barger and Phillips) is different from the Ln of Ni, and Ni, Lou, Lu, and Yang, and the P that I use is different from Pn, and the 2 M^2 that I use is ( 1 / 2 ) Sigma.

**Proposition 11.4 of chapter II of volume 1 of Kobayashi and Nomizu** states that

$$2FhII(X,Y) = [P(X),P(Y)] - P([X,Y])$$

where P takes values in the SU(2) Lie algebra. If the action of the Hodge dual * on P is such that *P = -P and *[P,P] = [P,P], then

$$FhII(X,Y) \wedge *FhII(X,Y) = (1/4) ( [ P(X) , P(Y) ]^2 - P([X,Y])^2 )$$

If integration of P over the Internal Symmetry Space gives P = (P+, P0), where P+ and P0 are the two components of the complex doublet scalar field, then

$$(1/4) ( [ P(X) , P(Y) ]^2 - P([X,Y])^2 ) = (1/4) ( L (PP)^2 - M^2 PP )$$

which is the Higgs Mechanism potential term.

In my notation (and that of Kane and Barger and Phillips), 2 M^2 is the square Mh^2 of the tree-level Higgs scalar particle mass.

In my notation (and that of Kane and Barger and Phillips), P is the Higgs scalar field, and its tree-level vacuum expectation value is given by

$$v^2 / 2 = P^2 = M^2 / 2 L \text{ or } M^2 = L v^2.$$

The value of the fundamental mass scale vacuum expectation value v of the Higgs scalar field is set in this model as the sum of the physical masses of the weak bosons, W+, W-, and Z0, whose tree-level masses will be 80.326 GeV, 80.326 GeV, and 91.862 GeV, respectively, and so that the electron mass will be 0.5110 MeV.

The resulting equations, in my notation (and that of Kane and Barger and Phillips), are:

$$Mh^2 = 2 M^2 \text{ and } M^2 = L v^2 \text{ and } Mh^2 / v^2 = 2 L$$

In their notation, Ni, Lou, Lu, and Yang have 2M^2 = (1/2) Sigma and P^2 = 6 Sigma / Ln, and for the tree-level value of the Higgs scalar particle mass Mh they have Mh^2 / Pn^2 = Ln / 3.

By combining the non-perturbative Gaussian Effective Potential (GEP) approach with their Regularization-Renormalization (R-R) method, Ni, Lou, Lu, and Yang find that:

Mh and Pn are the two fundamental mass scales of the Higgs mechanism, and the fundamental Higgs scalar field mass scale Pn of Ni, Lou, Lu, and Yang is equivalent to the vacuum expectation value v of the Higgs scalar field in my notation and that of Kane and Barger and Phillips, and

Ln (and the corresponding L) can not only be interpreted as the Higgs scalar field self-coupling constant, but also can be interpreted as determining the invariant ratio between the mass squares of the Higgs mechanism fundamental mass scales, Mh^2 and Pn^2 = v^2. Since the tree-level value of Ln is Ln = 1, and since Ln / 3 = Mh^2 / Pn^2 = Mh^2 / v^2 = 2 L, the tree-level value of L is L = Ln / 6 = 1 / 6, so that, at tree-level

Mh^2 / Pn^2 = Mh^2 / v^2 = 2 / 6 = 1 / 3.

In the theoretical model, the fundamental mass scale vacuum expectation value v of the Higgs scalar field is the fundamental mass parameter that is to be set to define all other masses by the mass ratio formulas of the model.

v is set to be 252.514 GeV

so that it is equal to the sum of the physical masses of the weak bosons, W+, W-, and Z0, whose tree-level masses will be 80.326 GeV, 80.326 GeV, and 91.862 GeV, respectively, and so that the electron mass will be 0.5110 MeV.

Then, the tree-level mass Mh of the Higgs scalar particle is given by

Mh = v / sqrt(3) = 145.789 GeV

The Higgs scalar field P is a Complex Doublet that can be expressed in terms of a vacuum expectation value v and a real Higgs field H.

The Complex Doublet P = ( P+, P0) = (1/sqrt(2)) ( P1 + iP2, P3 + iP4 ) = (1/sqrt(2)) ( 0, v + H ), so that

$$P3 = (1/\sqrt{2}) ( v + H )$$

where v is the vacuum expectation value and H is the real surviving Higgs field.

The value of the fundamental mass scale vacuum expectation value v of the Higgs scalar field is in the theoretical model set to be 252.514 GeV so that the electron mass will turn out to be 0.5110 MeV.

Now, to interpret the term

$$(1/4) ( [ P(X) , P(Y) ]^2 - P([X,Y])^2 ) = (1/4) ( L (PP)^2 - M^2 PP )$$

in terms of v and H, note that $L = M^2 / v^2$ and that $P = (1/\sqrt{2}) ( v + H )$, so that

$$FhII(X,Y) \wedge *FhII(X,Y) = (1/4) ( L (PP)^2 - M^2 PP ) =$$

$$= (1/16) ((M^2 / v^2) ( v + H )^4 - (1/8) M^2 ( v + H )^2 =$$

$$= (1/4) M^2 H^2 - (1/16) M^2 v^2 ( 1 - 4 H^3 / v^3 - H^4 / v^4 )$$

Disregarding some terms in v and H,

$$FhII(X,Y) \wedge *FhII(X,Y) = (1/4) M^2 H^2 - (1/16) M^2 v^2$$

**The second term** is the integral over 4-dim SpaceTime of the integral over 4-dim Internal Symmetry Space of

$$Fh4I \wedge *Fh4I$$

The second term after integration over the 4-dim Internal Symmetry Space, produces, by a process similar to the Mayer Mechanism, terms of the form

$$dP\ dP$$

where P is the Phi scalar complex doublet term and d is the covariant derivative.

**Proposition 11.4 of chapter II of volume 1 of Kobayashi and Nomizu** states that

$$2Fh4I(X,Y) = [P(X),P(Y)] - P([X,Y])$$

where P(X) takes values in the SU(2) Lie algebra. If the X component of Fh4I(X,Y) is in the surviving 4-dim SpaceTime and the Y component of Fh4I(X,Y) is in the 4-dim Internal Symmetry Space, then the Lie bracket product [X,Y] = 0 so that P([X,Y]) = 0 and therefore

$$Fh4I(X,Y) = (1/2) [P(X),P(Y)] = (1/2) dx P(Y)$$

Integration over Internal Symmetry Space of (1/2) dx P(Y) gives (1/2) dx P, where now P denotes the scalar Higgs field and dx denotes covariant derivative in the X direction.

Taking into account the Complex Doublet structure of P, the second term is the Integral over 4-dim SpaceTime of

$$Fh4I \wedge *Fh4I = (1/2) d P \wedge *(1/2) d P = (1/4) d P \wedge *d P =$$

$$= (1/4) (1/2) d ( v + H ) \wedge *d ( v + H ) = (1/8) dH dH + (\text{some terms in v and H})$$

Disregarding some terms in v and H,

$$Fh4I \wedge *Fh4I = (1/8) dH dH$$

Combining the second and third terms, since the first term is merged into the weak force part of the Lagrangian:

$$Fh4I \wedge *Fh4I + FhII(X,Y) \wedge *FhII(X,Y) =$$

$$= (1/8) dH dH + (1/4) M^2 H^2 - (1/16) M^2 v^2 =$$

$$= (1/8) ( dH dH + 2 M^2 H^2 - (1/2) M^2 v^2 )$$

This is the form of the Higgs Lagrangian in Barger and Phillips for a Higgs scalar particle of mass

$$Mh = M \, sqrt(2) = v / sqrt(3)= 145.789 \text{ GeV}$$

---

## To calculate Charge = Amplitude to Emit Gauge Boson and its probability-square, Force Strength:

```
Three factors determine the probability for emission of
a gauge boson from an origin spacetime vertex to a target vertex:

the part of the Internal Symmetry Space
of the target spacetime vertex that is available for the gauge boson
to go to from the origin vertex;

the volume of the spacetime link that is available for the gauge
boson to go through from the origin vertex to the target vertex; and

an effective mass factor for forces
(such as the Weak force and Gravity)
that, in the low-energy ranges of our experiments,
are carried effectively by gauge bosons that are not
massless high-energy.

In this physics model, force strength probabilities are calculated
in terms of relative volumes of bounded complex homogeneous domains and
their Shilov boundaries.

The bounded complex homogeneous domains correspond to
harmonic functions of generalized Laplacians
that determine heat equations, or diffusion equations;

while the amplitude to emit gauge bosons in the
HyperDiamond Feynman Checkerboard is a process that
is similar to diffusion, and
therefore also corresponds to a generalized Laplacian.


In this theoretical  model, all force strengths are
represented as ratios with respect to the geometric force
strength of Gravity (that is, the force strength of Gravity
without using the Effective Mass factor).
```

Therefore, the only free charge, or force strength, parameter is the charge of the Spin(5) gravitons in the MacDowell-Mansouri formalism of Gravity. Note that these Spin(5) gravitons are NOT the ordinary spin-2 gravitons of the low-energy region in which we live.
The charge of the Spin(5) gravitons is taken to be unity, 1, so that its force strength is also unity, 1.
All other force strengths are determined as ratios with respect to the Spin(5) gravitons and each other.

The force strength probability for a gauge boson to be emitted from an origin spacetime HyperDiamond vertex and go to a target vertex is the product of three things:

the volume Vol(MISforce) of the target Internal Symmetry Space, that is, the part of the Internal Symmetry Space of the target spacetime vertex that is available for the gauge boson to go to from the origin vertex;

the volume Vol(Qforce) / Vol(Dforce)^( 1 / mforce ) of the spacetime link to the target spacetime vertex from the origin vertex; and an effective mass factor 1 / Mforce^2 for forces (such as the Weak force and Gravity) that, in the low-energy ranges of our experiments, are carried effectively by gauge bosons that are not massless high-energy SU(2) or Spin(5) gauge bosons, but are either massive Weak bosons due to the Higgs mechanism or effective spin-2 gravitons. For other forces, the effective mass factor is taken to be unity, 1.

Therefore, the force strength of a given force is

```
alphaforce = (1 / Mforce^2 \)
             ( Vol(MISforce))
             ( Vol(Qforce) / Vol(Dforce)^( 1 / mforce ))
```

where:

alphaforce represents the force strength;

Mforce represents the effective mass;

MISforce represents the part of the target
Internal Symmetry Space that is available for the gauge
boson to go to;

Vol(MISforce) stands for volume of MISforce,
and is sometimes also denoted by the shorter notation Vol(M);

Qforce represents the link from the origin
to the target that is available for the gauge
boson to go through;

Vol(Qforce) stands for volume of Qforce;

Dforce represents the complex bounded homogeneous domain
of which Qforce is the Shilov boundary;

mforce is the dimensionality of Qforce,
which is 4 for Gravity and the Color force,
2 for the Weak force (which therefore is considered to
have two copies of QW for each spacetime HyperDiamond link),
and 1 for Electromagnetism (which therefore is considered to
have four copies of QE for each spacetime HyperDiamond link)

Vol(Dforce)^( 1 / mforce ) stands for
a dimensional normalization factor (to reconcile the dimensionality
of the Internal Symmetry Space of the target vertex
with the dimensionality of the link from the origin to the
target vertex).

The Qforce, Hermitian symmetric space,
and Dforce manifolds for the four forces are:

| Gauge Group | Hermitian Symmetric Space | Type of Dforce | mforce | Qforce |
|---|---|---|---|---|
| Spin(5) | Spin(7) / Spin(5)xU(1) | IV5 | 4 | RP^1xS^4 |
| SU(3) | SU(4) / SU(3)xU(1) | B^6(ball) | 4 | S^5 |
| SU(2) | Spin(5) / SU(2)xU(1) | IV3 | 2 | RP^1xS^2 |
| U(1) | - | - | 1 | - |

The geometric volumes needed for the calculations, mostly taken from Hua, are

| Force | M | Vol(M) | Q | Vol(Q) | D | Vol(D) |
|---|---|---|---|---|---|---|
| gravity | S^4 | 8pi^2/3 | RP^1xS^4 | 8pi^3/3 | IV5 | pi^5/2^4 5! |
| color | CP^2 | 8pi^2/3 | S^5 | 4pi^3 | B^6(ball) | pi^3/6 |
| weak | S^2xS^2 | 2x4pi | RP^1xS^2 | 4pi^2 | IV3 | pi^3/24 |
| e-mag | T^4 | 4x2pi | - | - | - | - |

Using these numbers, the results of the calculations are the relative force strengths at the characteristic energy level of the generalized Bohr radius of each force:

| Gauge Group | Force | Characteristic Energy | Geometric Force Strength | Total Force Strength |
|---|---|---|---|---|
| Spin(5) | gravity | approx 10^19 GeV | 1 | GGmproton^2 approx 5 x 10^-39 |
| SU(3) | color | approx 245 MeV | 0.6286 | 0.6286 |
| SU(2) | weak | approx 100 GeV | 0.2535 | GWmproton^2 approx 1.05 x 10^-5 |
| U(1) | e-mag | approx 4 KeV | 1/137.03608 | 1/137.03608 |

The force strengths are given at the characteristic energy levels of their forces, because the force strengths run with changing energy levels.

The effect is particularly pronounced with the color force.

The color force strength was calculated using a simple perturbative QCD renormalization group equation at various energies, with the following results:

| Energy Level | Color Force Strength |
|---|---|
| 245 MeV | 0.6286 |
| 5.3 GeV | 0.166 |
| 34 GeV | 0.121 |
| 91 GeV | 0.106 |

Taking other effects, such as Nonperturbative QCD, into account, should give a Color Force Strength of about 0.125 at about 91 GeV

---

## To calculate Weak Boson Masses and Weinberg Angle:

Denote the 3 SU(2) high-energy weak bosons (massless at energies higher than the electroweak unification) by W+, W-, and W0, corresponding to the massive physical weak bosons W+, W-, and Z0.

The triplet { W+, W-, W0 } couples directly with the T - Tbar quark-antiquark pair, so that the total mass of the triplet { W+, W-, W0 } at the electroweak unification is equal to the total mass of a T - Tbar pair, 259.031 GeV.

```
The triplet  { W+, W-, Z0 }
couples directly with the Higgs scalar,
which carries the Higgs mechanism by
which the W0 becomes the physical Z0,
so that the total mass of the triplet  { W+, W-, Z0 }
is equal to the vacuum expectation value v of
the Higgs scalar field, v = 252.514 GeV.

What are individual masses of members
of the triplet { W+, W-, Z0 } ?

First, look at the triplet  { W+, W-, W0 }
which can be represented by the 3-sphere S^3.

The Hopf fibration of S^3 as
S^1 --} S^3 --} S^2
gives a decomposition of the W bosons
into the neutral W0 corresponding to S^1 and
the charged pair W+ and W- corresponding
to S^2.


The mass ratio of the sum of the masses of
W+ and W- to
the mass of W0
should be the volume ratio of
the S^2 in S^3 to
the S^1 in S3.


The unit sphere S^3 in R^4 is
normalized by 1 / 2.


The unit sphere S^2 in R^3 is
normalized by 1 / sqrt3.


The unit sphere S^1 in R^2 is
normalized by 1 / sqrt2.


The ratio of the sum of the W+ and W- masses to
```

the W0 mass should then be
(2 / sqrt3) V(S^2) / (2 / sqrt2) V(S^1) = 1.632993.

Since the total mass of the triplet { W+, W-, W0 }
is 259.031 GeV, the total mass of a T - Tbar pair,
and the charged weak bosons have equal mass,
we have

mW+ = mW- = 80.326 GeV,

and mW0 = 98.379 GeV.

## Parity Violation, Effective Masses, and Weinberg Angle:

The charged W+/- neutrino-electron interchange must be symmetric with the electron-neutrino interchange,

so that the absence of right-handed neutrino particles requires

that the charged W+/- SU(2)
weak bosons act only on left-handed electrons.

Each gauge boson must act consistently
on the entire Dirac fermion particle sector,
so that the charged W+/- SU(2) weak bosons
act only on left-handed fermions of all types.

The neutral W0 weak boson does not interchange Weyl
neutrinos with Dirac fermions, and so is not restricted
to left-handed fermions,
but also has a component that acts on both types of fermions,
both left-handed and right-handed, conserving parity.

However, the neutral W0 weak bosons are related to
the charged W+/- weak bosons by custodial SU(2)
symmetry, so that the left-handed component of the
neutral W0 must be equal to the left-handed (entire)
component of the charged W+/-.

Since the mass of the W0 is greater than the mass
of the W+/-, there remains for the W0 a component
acting on both types of fermions.

Therefore the full W0 neutral weak boson interaction
is proportional to
(mW+/-^2 / mW0^2) acting on left-handed fermions
and

(1 - (mW+/-^2 / mW0^2)) acting
on both types of fermions.

If (1 - (mW+/-2 / mW0^2)) is defined to be
sin(thetaw)^2 and denoted by K, and

if the strength of the W+/- charged weak force
(and of the custodial SU(2) symmetry) is denoted by T,

then the W0 neutral weak interaction can be written as:

W0L = T + K and W0LR = K.

Since the W0 acts as W0L with respect to the
parity violating SU(2) weak force and

as W0LR with respect to the parity conserving U(1)
electromagnetic force of the U(1) subgroup of SU(2),

the W0 mass mW0 has two components:

the parity violating SU(2) part mW0L that is
equal to mW+/- ; and

the parity conserving part mW0LR that acts like a
heavy photon.

As mW0 = 98.379 GeV = mW0L + mW0LR, and

as mW0L = mW+/- = 80.326  GeV,

we have mW0LR = 18.053  GeV.

Denote by *alphaE = *e^2 the force
strength of the weak parity conserving U(1)
electromagnetic type force that acts through the
U(1) subgroup of SU(2).

The electromagnetic force strength
alphaE = e^2 = 1 / 137.03608 was calculated
in Chapter 8 using
the volume V(S^1) of an S^1 in R^2,
normalized by 1 / \qrt2.

The *alphaE force is part of the SU(2) weak
force whose strength alphaW = w^2 was calculated
in Chapter 8 using the volume V(S^2) of an S^2 \subset R^3,
normalized by 1 / sqrt3.

Also, the electromagnetic force strength alphaE = e^2
was calculated in Chapter 8 using a
4-dimensional spacetime with global structure of
the 4-torus T^4 made up of four S^1 1-spheres,

while the SU(2) weak force strength
alphaW = w^2 was calculated in Chapter 8
using two 2-spheres S^2 x S^2,
each of which contains one 1-sphere of
the *alphaE force.

Therefore
*alphaE = alphaE ( sqrt2 / sqrt3)(2 / 4) = alphaE / sqrt6,

*e = e / (4th root of 6) = e / 1.565 , and

the mass mW0LR must be reduced to an effective value

mW0LReff = mW0LR / 1.565 = 18.053/1.565 = 11.536 GeV

for the *alphaE force to act like
an electromagnetic force in the 4-dimensional
spacetime HyperDiamond Feynman Checkerboard model:

*e mW0LR = e (1/5.65) mW0LR = e mZ0,

where the physical effective neutral weak boson is
denoted by Z0.

Therefore, the correct HyperDiamond Feynman Checkerboard values for
weak boson masses and the Weinberg angle thetaW are:

mW+ = mW- = 80.326 GeV;

mZ0 = 80.326 + 11.536 = 91.862 GeV;

Sin(thetaW)^2 = 1 - (mW+/- / mZ0)^2 =

= 1 - ( 6452.2663 / 8438.6270 ) = 0.235.

Radiative corrections are not taken into account here,
and may change these tree-level values somewhat.

---

## To calculate Fermion Masses:

### Constituent Quark Masses:

To do calculations in theories such as
Perturbative QCD and Chiral Perturbation Theory,
you need to use effective quark masses that
are called current masses.  Current quark masses are different

```
from the Pre-Quantum constituent quark masses of our model.
The current mass of a quark is defined in this model as
the difference between
the constituent mass of the quark
and
the density of the lowest-energy sea of virtual gluons,
quarks, and antiquarks, or 312.75 MeV.
```

The fundamental correctness of the Constituent Quark Mass and the effectiveness of the NonRelativistic Quark Model of hadrons can be explained by Bohm's quantum theory applied to a fermion confined in a box, in which the fermion is at rest because its kinetic energy is transformed into PSI-field potential energy. Since that aspect of Bohm's quantum theory is not a property of most other formulations of quantum theory, the effectiveness of the NonRelativistic Quark Model confirms Bohm's quantum theory as opposed to those others.

Fermion Mass Calculations:

```
First generation fermion particles are also represented
by octonions as follows:



  Octonion                  Fermion
Basis Element               Particle

     1                     e-neutrino

     i                    red  up  quark
     j                  green  up  quark
     k                   blue  up  quark

     E                      electron

     I                   red  down  quark
     J                 green  down  quark
     K                  blue  down  quark



First generation fermion antiparticles are represented
```

by octonions in a similiar way.

Second generation fermion particles and antiparticles
are represented by pairs of octonions.

Third generation fermion particles and antiparticles
are represented by triples of octonions.

There are no higher generations of fermions than the Third.

This can be seen geometrically as a consequence of the fact
that
if you reduce the original 8-dimensional spacetime
into associative 4-dimensional physical spacetime
and coassociative 4-dimensional Internal Symmetry Space
then
if you look in the original 8-dimensional spacetime
at a fermion (First-generation represented by a single octonion)
propagating from one vertex to another
there are only 4 possibilities for the same propagation
after dimensional reduction:

1 - the origin o and target x vertices are both
in the associative 4-dimensional physical spacetime

```
4-dim Internal Symmetry Space   --------------


4-dim Physical SpaceTime        ---o------x---
```

in which case the propagation is unchanged, and the
fermion remains a FIRST generation fermion represented
by a single octonion o

2 - the origin vertex o is in the associative spacetime
and the target vertex * in in the Internal Symmetry Space

```
4-dim Internal Symmetry Space   ----------*---


4-dim Physical SpaceTime        ---o----------
```

in which case there must be a new link from
the original target vertex * in the Internal Symmetry Space

```
to a new target vertex x in the associative spacetime

4-dim Internal Symmetry Space   ----------*---


4-dim Physical SpaceTime        ---o------x---

   and a second octonion can be introduced at the original
target vertex in connection with the new link
so that the fermion can be regarded after dimensional reduction
as a pair of octonions o and *
and therefore as a SECOND generation fermion



3 - the target vertex x is in the associative spacetime
and the origin vertex o in in the Internal Symmetry Space

4-dim Internal Symmetry Space   ---o----------


4-dim Physical SpaceTime        ----------x---

   in which case there must be a new link to
the original origin vertex o in the Internal Symmetry Space
from a new origin vertex * in the associative spacetime

4-dim Internal Symmetry Space   ---o----------


4-dim Physical SpaceTime        ---O------x---

   so that a second octonion can be introduced at the new
origin vertex O in connection with the new link
so that the fermion can be regarded after dimensional reduction
as a pair of octonions O and o
and therefore as a SECOND generation fermion


and


4 - both the origin vertex o and the target vertex *
are in the Internal Symmetry Space,

4-dim Internal Symmetry Space   ---o------*---
```

```
4-dim Physical SpaceTime        --------------

   in which case there must be a new link to
the original origin vertex o in the Internal Symmetry Space
from a new origin vertex O in the associative spacetime,
and a second new link from the original target vertex *
in the Internal Symmetry Space to a new target vertex x
in the associative spacetime

4-dim Internal Symmetry Space   ---o------*---


4-dim Physical SpaceTime        ---O------x---
```

so that a second octonion can be introduced at the new origin vertex O in connection with the first new link, and a third octonion can be introduced at the original target vertex * in connection with the second new link, so that the fermion can be regarded after dimensional reduction as a triple of octonions O and o and * and therefore as a THIRD generation fermion.

As there are no more possibilities, there are no more generations, and we have:

The first generation fermions
correspond to octonions   O

and second generation fermions
correspond to pairs of octonions   O x  O

and third generation fermions
correspond to triples of octonions   O x  O x  O

To calculate the fermion masses in the model, the volume of a compact manifold representing the spinor fermions S8+ is used. It is the parallelizable manifold S^7 x RP^1.

Also, since gravitation is coupled to mass, the infinitesimal generators of the MacDowell-Mansouri gravitation group, Spin(0,5), are relevant.

The calculated quark masses are constituent masses,
not current masses.

Fermion masses are calculated as a product of four factors:

V(Qfermion) x N(Graviton) x N(octonion) x Sym

V(Qfermion) is the volume of the part of
the half-spinor fermion particle manifold S^7 x RP^1
that is related to the fermion particle
by photon, weak boson, and gluon interactions.

N(Graviton) is the number of types of Spin(0,5) graviton
related to the fermion.
The 10 gravitons correspond to
the 10 infinitesimal generators of Spin(0,5) = Sp(2).

2 of them are in the Cartan subalgebra.

6 of them carry color charge,
and may therefore be considered as corresponding to quarks.

The remaining 2 carry no color charge,
but may carry electric charge
and so may be considered as corresponding to electrons.

One graviton takes the electron into itself,
and the other can only take the first-generation electron
into the massless electron neutrino.

Therefore only one graviton should correspond to the mass
of the first-generation electron.

The graviton number ratio of the down quark to the first-generation electron is therefore 6/1 = 6.

N(octonion) is an octonion number factor relating up-type quark masses to down-type quark masses in each generation.

Sym is an internal symmetry factor, relating 2nd and 3rd generation massive leptons to first generation fermions. It is not used in first-generation calculations.

The ratio of the down quark constituent mass to the electron mass is then calculated as follows:

Consider the electron, e.

By photon, weak boson, and gluon interactions,
e can only be taken into 1, the massless neutrino.

The electron and neutrino, or their antiparticles,
cannot be combined to produce any of the
massive up or down quarks.

The neutrino, being massless at tree level,
does not add anything to the mass formula for the electron.

Since the electron cannot be related to any other massive Dirac fermion, its volume V(Qelectron) is taken to be 1.

Next consider a red down quark I.

By gluon interactions,
I can be taken into J and K,
the blue and green down quarks.

By also using weak boson interactions,
it can be taken into i, j, and k,
the red, blue, and green up quarks.

Given the up and down quarks,
pions can be formed from quark-antiquark pairs,
and the pions can decay

to produce electrons and neutrinos.

Therefore the red down quark (similarly, any down quark)
is related to any part of S^7 x RP^1,
the compact manifold corresponding to

{ 1, i, j, k, I, J, K, E }

and therefore a down quark should have a spinor manifold
volume factor V(Qdown quark) of the volume of
S^7 x RP^1.

The ratio of the down quark spinor manifold volume factor to
the electron spinor manifold volume factor is just

V(Qdown quark) / V(Qelectron) = V(S^7x  RP^1)/1 = pi^5 / 3.

Since the first generation graviton factor is 6,

md/me = 6V(S^7 x RP^1) = 2 pi^5 = 612.03937

As the up quarks correspond to i, j, and k,
which are
the octonion transforms under E of I, J, and K
of the down quarks,
the up quarks and down quarks
have the same constituent mass mu = md.

Antiparticles have the same mass as the corresponding
particles.

Since the model only gives ratios of massses,
the mass scale is fixed so that the electron mass

me = 0.5110 MeV.

Then, the constituent mass of the down quark is

md = 312.75 MeV, and

the constituent mass for the up quark is

mu = 312.75 MeV.

These results when added up give a total mass of
first generation fermion particles:

Sigmaf1 = 1.877 GeV

As the proton mass is taken to be the sum of the constituent masses of its constituent quarks

mproton = mu + mu + md = 938.25 MeV

The theoretical calculation is close to
the experimental value of 938.27 MeV.

The third generation fermion particles correspond to triples of octonions. There are 8^3 = 512 such triples.

The triple { 1,1,1 } corresponds to the tau-neutrino.

The other 7 triples involving only 1 and E correspond
to the tauon:
{ E, E, E }
{ E, E, 1 }
{ E, 1, E }
{ 1, E, E }
{ 1, 1, E }
{ 1, E, 1 }
{ E, 1, 1 }

The symmetry of the 7 tauon triples is the same
as the symmetry of
the 3 down quarks, the 3 up quarks, and the electron,
so the tauon mass should be

the same as
the sum of the masses of
the first generation massive fermion particles.

## Therefore the tauon mass is 1.877 GeV.

The calculated Tauon mass of 1.88 GeV is a sum
of first generation fermion masses, all of which are
valid at the energy level of about 1 GeV.

However, as the Tauon mass is about 2 GeV,
the effective Tauon mass should be renormalized
from the energy level of 1 GeV (where the mass is 1.88 GeV)
to the energy level of 2 GeV.
Such a renormalization should reduce the mass.
If the renormalization reduction were about 5 percent,
the effective Tauon mass at 2 GeV would be about 1.78 GeV.

The 1996 Particle Data Group Review of Particle Physics gives
a Tauon mass of 1.777 GeV.

---

Note that all triples corresponding to the
tau and the tau-neutrino are colorless.

The beauty quark corresponds to 21 triples.

They are triples of the same form as the 7 tauon triples,
but for 1 and I, 1 and J, and 1 and K,
which correspond to the red, green, and blue beauty quarks,
respectively.

The seven triples of the red beauty quark correspond
to the seven triples of the tauon,
except that the beauty quark interacts with 6 Spin(0,5)
gravitons while the tauon interacts with only two.

The beauty quark constituent mass should be the tauon mass times the third generation graviton factor 6/2 = 3, so the B-quark mass is

$$mb = 5.63111 \text{ GeV}.$$

The calculated Beauty Quark mass of 5.63 GeV is a consitituent mass, that is,
it corresponds to the conventional pole mass plus 312.8 MeV.

Therefore, the calculated Beauty Quark mass of 5.63 GeV corresponds to a conventional pole mass of 5.32 GeV.

The 1996 Particle Data Group Review of Particle Physics gives a lattice gauge theory Beauty Quark pole mass as 5.0 GeV.

The pole mass can be converted to an MSbar mass
if the color force strength constant alpha_s is known.
The conventional value of alpha_s at about 5 GeV is about 0.22.
Using alpha_s (5 GeV) = 0.22,
a pole mass of 5.0 GeV gives an MSbar 1-loop mass of 4.6 GeV,
and an MSbar 1,2-loop mass of 4.3, evaluated at about 5 GeV.

If the MSbar mass is run from 5 GeV up to 90 GeV,
the MSbar mass decreases by about 1.3 GeV,
giving an expected MSbar mass of about 3.0 GeV at 90 GeV.
DELPHI at LEP has observed the Beauty Quark
and found a 90 GeV MSbar mass of about 2.67 GeV,
with error bars +/- 0.25 (stat) +/- 0.34 (frag) +/- 0.27 (theo).

Note that the theoretical model calculated mass of 5.63 GeV
corresponds to a pole mass of 5.32 GeV,
which is somewhat higher than the conventional value of 5.0 GeV.
However,
the theoretical model calculated value of
the color force strength constant alpha_s at
about 5 GeV is about 0.166,
while
the conventional value of
the color force strength constant alpha_s at about 5 GeV is about 0.216,
and
the theoretical model calculated value of
the color force strength constant alpha_s at
about 90 GeV is about 0.106,
while
the conventional value of
the color force strength constant alpha_s at
about 90 GeV is about 0.118.

The theoretical model calculations gives
a Beauty Quark pole mass (5.3 GeV)
that is about 6 percent higher
than the conventional Beauty Quark pole mass (5.0 GeV),
and
a color force strength alpha_s at 5 GeV (0.166) such that
1 + alpha_s = 1.166 is about 4 percent lower
than the conventional value of 1 + alpha_s = 1.216 at 5 GeV.

---

Note particularly that triples of the type { 1, I, J },
{ I, J, K }, etc.,
do not correspond to the beauty quark, but to the truth quark.

The truth quark corresponds to the remaining 483 triples, so the
constituent mass of the red truth quark is 161/7 = 23 times the
red beauty quark mass, and the red T-quark mass is

$$mt = 129.5155 \text{ GeV}$$

The blue and green truth quarks are defined similarly.

All other masses than the electron mass
(which is the basis of the assumption of the value of the
Higgs scalar field vacuum expectation value v = 252.514 GeV),
including the Higgs scalar mass and Truth quark mass,
are calculated (not assumed) masses in the HyperDiamond Feynman
Checkerboard model.

The tree level T-quark constituent mass rounds off to 130 GeV.

These results when added up give a total mass of
third generation fermion particles:

Sigmaf3 = 1,629 GeV

The second generation fermion calculations are:

The second generation fermion particles correspond
to pairs of octonions.

There are 8^2 = 64 such pairs.

The pair { 1,1 } corresponds to the mu-neutrino.

the pairs { 1, E }, { E, 1 }, and
{ E, E } correspond to the muon.

Compare the symmetries of the muon pairs to the symmetries
of the first generation fermion particles.

The pair { E, E } should correspond
to the E electron.

The other two muon pairs have a symmetry group S2,
which is 1/3 the size of the color symmetry group S3
which gives the up and down quarks their mass of 312.75 MeV.

Therefore the mass of the muon should be the sum of
the { E, E } electron mass and
the { 1, E }, { E, 1 } symmetry mass,
which is 1/3 of the up or down quark mass.

Therefore, mmu = 104.76 MeV.

According to the 1998 Review of Particle Physics of the Particle Data Group, the experimental muon mass is about 105.66 MeV.

```
Note that all pairs corresponding to
the muon and the mu-neutrino are colorless.



The red, blue and green strange quark each corresponds
to the 3 pairs involving 1 and I, J, or K.


The red strange quark is defined as the thrge pairs
1 and I, because I is the red down quark.

Its mass should be the sum of two parts:
the { I, I } red down quark mass, 312.75 MeV, and
the product of the symmetry part of the muon mass, 104.25 MeV,
times the graviton factor.


Unlike the first generation situation,
massive second and third generation leptons can be taken,
by both of the colorless gravitons that
may carry electric charge, into massive particles.


Therefore the graviton factor for the second and

third generations is 6/2 = 3.


Therefore the symmetry part of the muon mass times
the graviton factor 3 is 312.75 MeV,
and
the red strange quark constituent mass
is
```

$$ms = 312.75 \text{ MeV} + 312.75 \text{ MeV} = 625.5 \text{ MeV}$$

```
The blue strange quarks correspond to the
three pairs involving J,
the green strange quarks correspond to the
```

three pairs involving K,
and their masses are determined similarly.

The charm quark corresponds to the other 51 pairs.
Therefore, the mass of the red charm quark should
be the sum of two parts:

the { i, i }, red up quark mass, 312.75 MeV; and

the product of the symmetry part of the strange quark
mass, 312.75 MeV, and

the charm to strange octonion number factor 51/9,
which product is 1,772.25 MeV.

Therefore the red charm quark constituent mass
is

$$mc = 312.75 \text{ MeV} + 1{,}772.25 \text{ MeV} = 2.085 \text{ GeV}$$

The blue and green charm quarks are defined similarly,
and their masses are calculated similarly.

The calculated Charm Quark mass of 2.09 GeV is a
consitituent mass, that is,
it corresponds to the conventional pole mass plus 312.8 MeV.

Therefore, the calculated Charm Quark mass of 2.09 GeV
corresponds to a conventional pole mass of 1.78 GeV.

The 1996 Particle Data Group Review of Particle Physics gives
a range for the Charm Quark pole mass from 1.2 to 1.9 GeV.

The pole mass can be converted to an MSbar mass
if the color force strength constant alpha_s is known.
The conventional value of alpha_s at about 2 GeV is about 0.39,

which is somewhat lower than the teoretical model value.
Using alpha_s (2 GeV) = 0.39,
a pole mass of 1.9 GeV gives an MSbar 1-loop mass of 1.6 GeV,
evaluated at about 2 GeV.

These results when added up give a total mass of
second generation fermion particles:

Sigmaf2 = 32.9 GeV

---

## To calculate Kobayashi-Maskawa Parameters:

The Kobayashi-Maskawa parameters are determined in terms of
the sum of the masses of the 30 first-generation
fermion particles and antiparticles, denoted by Smf1 = 7.508 GeV,
and
the similar sums for second-generation and third-generation fermions,
denoted by Smf2 = 32.94504 GeV and Smf3 = 1,629.2675 GeV.

The reason for using sums of all fermion masses
(rather than sums of quark masses only) is that
all fermions are in the same spinor representation of Spin(8),
and the Spin(8) representations are considered to be fundamental.

The following formulas use the above masses to
calculate Kobayashi-Maskawa parameters:

phase angle d13 = 1 radian ( unit length on a phase circumference )

```
sin(alpha) = s12 =
= [me+3md+3mu]/sqrt([me^2+3md^2+3mu^2]+[mmu^2+3ms^2+3mc^2]) =
          = 0.222198

sin(beta) = s13 =
```

```
= [me+3md+3mu]/sqrt([me^2+3md^2+3mu^2]+[mtau^2+3mb^2+3mt^2])=
          = 0.004608

sin(*gamma) =
= [mmu+3ms+3mc]/sqrt([mtau^2+3mb^2+3mt^2]+[mmu^2+3ms^2+3mc^2])

sin(gamma) = s23 = sin(*gamma) sqrt( Sigmaf2 / Sigmaf1 ) =
           = 0.04234886

The factor sqrt( Smf2 /Smf1 ) appears in s23 because
an s23 transition is to the second generation and
not all the way to the first generation,
so that the end product of an s23 transition has a
greater available energy than s12 or s13 transitions
by a factor of Smf2 / Smf1 .
Since the width of a transition is proportional to
the square of the modulus of the relevant KM entry and
the width of an s23 transition has greater available energy
than the s12 or s13 transitions by a factor of Smf2 / Smf1
the effective magnitude of the s23 terms in the KM entries
is increased by the factor sqrt( Smf2 /Smf1 ) .
```

The Chau-Keung parameterization is used,
as it allows the K-M matrix to be represented as
the product of the following three 3x3 matrices:

---

| | | |
|---|---|---|
| 1 | 0 | 0 |
| 0 | cos(gamma) | sin(gamma) |
| 0 | -sin(gamma) | cos(gamma) |

---

| | | |
|---|---|---|
| cos(beta) | 0 | sin(beta)exp(-i d13) |
| 0 | 1 | 0 |
| -sin(beta)exp(i d13) | 0 | cos(beta) |

---

| | | |
|---|---|---|
| cos(alpha) | sin(alpha) | 0 |
| -sin(alpha) | cos(alpha) | 0 |
| 0 | 0 | 1 |

---

The resulting Kobayashi-Maskawa parameters
for W+ and W- charged weak boson processes, are:

| | d | s | b |
|---|---|---|---|
| u | 0.975 | 0.222 | 0.00249 -0.00388i |
| c | -0.222 -0.000161i | 0.974 -0.0000365i | 0.0423 |
| t | 0.00698 -0.00378i | -0.0418 -0.00086i | 0.999 |

The matrix is labelled by either
(u c t) input and (d s b) output,
or, as above,
(d s b) input and (u c t) output.

For Z0 neutral weak boson processes, which are suppressed
by the GIM mechanism of cancellation of virtual subprocesses,
the matrix is labelled by either
(u c t) input and (u'c't') output,
or, as below,
(d s b) input and (d's'b') output:

| | d | s | b |
|---|---|---|---|
| d' | 0.975 | 0.222 | 0.00249 -0.00388i |
| s' | -0.222 -0.000161i | 0.974 -0.0000365i | 0.0423 |
| b' | 0.00698 -0.00378i | -0.0418 -0.00086i | 0.999 |

Since neutrinos of all three generations are massless at tree level, the lepton sector has no tree-level K-M mixing.

According to a Review on the KM mixing matrix by Gilman, Kleinknecht, and Renk in the 2002 Review of Particle Physics:

"... Using the eight tree-level constraints discussed below together with unitarity, and assuming only three generations, the 90% confidence limits on the magnitude of the elements of the complete matrix are

| | d | s | b |
|---|---|---|---|
| u | 0.9741 to 0.9756 | 0.219 to 0.226 | 0.00425 to 0.0048 |
| c | 0.219 to 0.226 | 0.9732 to 0.9748 | 0.038 to 0.044 |
| t | 0.004 to 0.014 | 0.037 to 0.044 | 0.9990 to 0.9993 |

... The constraints of unitarity connect different elements, so choosing a specific value for one element restricts the range of others. ... The phase d13 lies in the range 0 < d13 < 2 pi, with non-zero values generally breaking CP invariance for the weak interactions. ... Using tree-level processes as constraints only, the matrix elements ...[ of the 90% confidence limit shown above ]... correspond to values of the sines of the angles of s12 = 0.2229 +/- 0.0022, s23 = 0.0412 +/- 0.0020, and s13 = 0.0036 +/- 0.0007. If we use the loop-level processes discussed below as additional constraints, the sines of the angles remain unaffected, and the CKM phase, sometimes referred to as the angle gamma = phi3 of the unitarity triangle ...

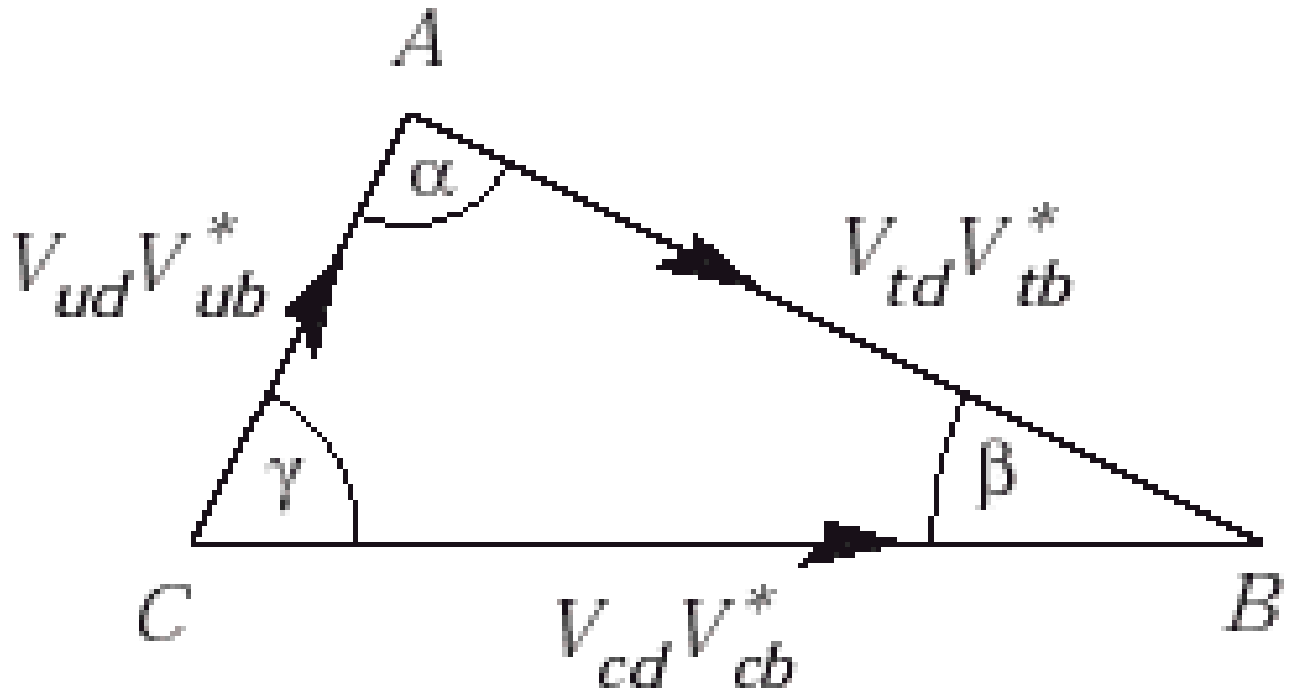

... is restricted to d13 = ( 1.02 +/- 0.22 ) radians = 59 +/- 13 degrees. ... CP-violating amplitudes or differences of rates are all proportional to the product of CKM factors ... s12 s13 s23 c12 c13^2 c23 sind13. This is just twice the area of the unitarity triangle. ... All processes can be quantitatively understood by one value of the CKM phase d13 = 59 +/- 13 degrees. The value of beta = 24 +/- 4 degrees from the overall fit is consistent with the value from the CP-asymmetry measurements of 26 +/- 4 degrees. The invariant measure of CP violation is J = ( 3.0 +/- 0.3) x 10^(-5). ... From a combined fit using the direct measurements, B mixing, epsilon, and sin2beta, we obtain: Re Vtd = 0.0071 +/- 0.0008 , Im Vtd = -0.0032 +/- 0.0004 ... Constraints... on the position of the apex of the unitarity triangle following from | Vub | , B mixing, epsilon, and sin2beta. ...

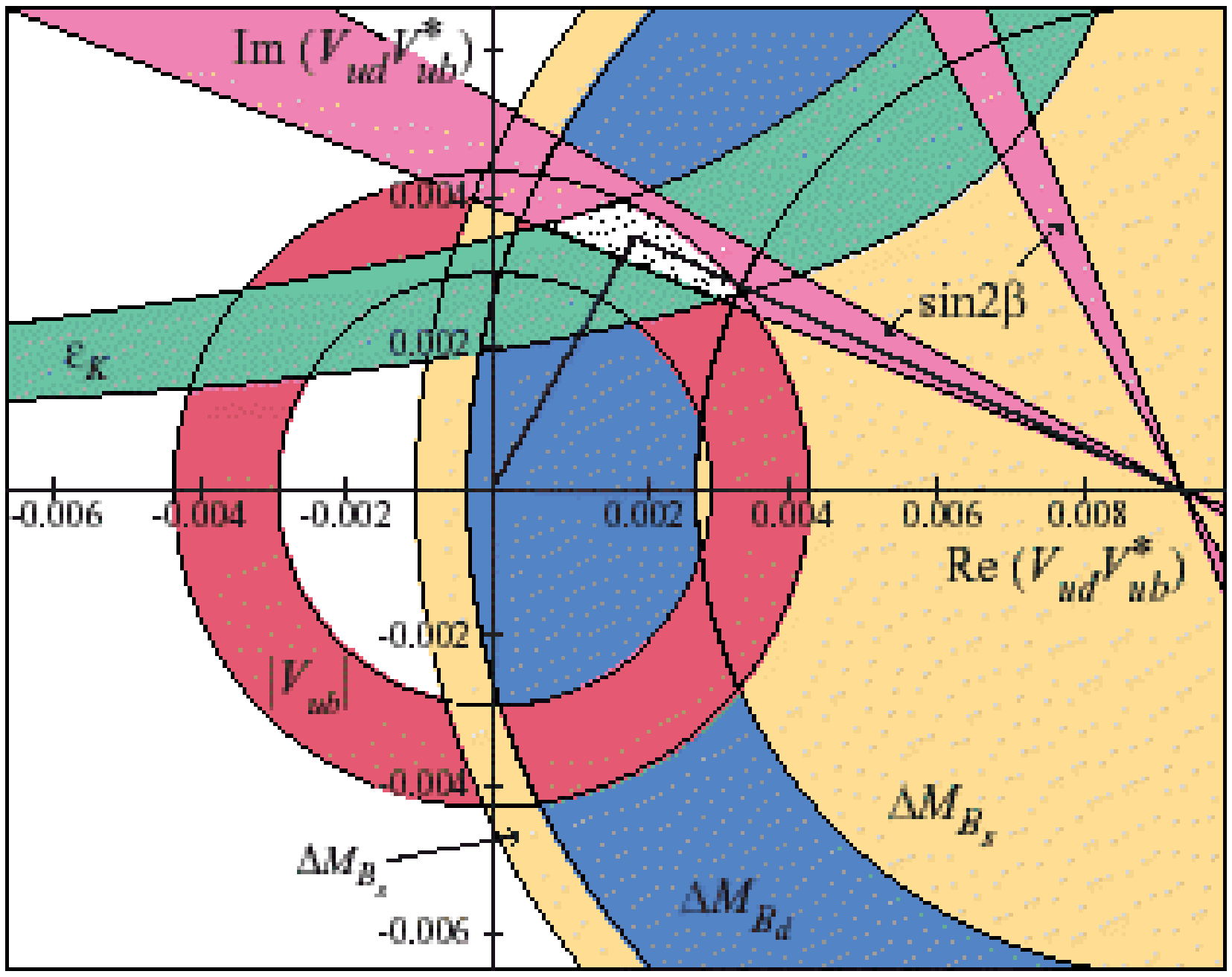

... A possible unitarity triangle is shown with the apex in the preferred region. ...".

In hep-ph/0208080, Yosef Nir says: "... Within the Standard Model, the only source of CP violation is the Kobayashi-Maskawa (KM) phase ... The study of CP violation is, at last, experiment driven. ... The CKM matrix provides a consistent picture of all the measured flavor and CP violating processes. ... There is no signal of new flavor physics. ... Very likely, the KM mechanism is the dominant source of CP violation in flavor changing processes. ... The result is consistent with the SM predictions. ...".

---

## Proton-Neutron Mass Difference:

```
According to the 1986 CODATA Bulletin No. 63,
the experimental value of the neutron mass is 939.56563(28) Mev,
```

and the experimental value of the proton is 938.27231(28) Mev.

The neutron-proton mass difference 1.3 Mev is due to
the fact that the proton consists of two up quarks and one down quark,
while the neutron consists of one up quark and two down quarks.

**The magnitude of the electromagnetic energy difference**
**mN - mP is about 1 Mev,**
**but the sign is wrong: mN - mP = -1 Mev,**
and
the proton's electromagnetic mass is greater than the neutron's.

The difference in energy between the bound states, neutron and proton,
is not due to a difference between
the Pre-Quantum constituent masses of the up quark and the down quark,
calculated in the theory to be equal.

It is due to the difference between the Quantum color force
interactions of the up and down constituent valence quarks
with the gluons and virtual sea quarks in the neutron and the proton.

An up valence quark, constituent mass 313 Mev,
does not often swap places with a 2.09 Gev charm sea quark,
but a 313 Mev down valence quark can more often swap places
with a 625 Mev strange sea quark.

Therefore the Quantum color force constituent mass
of the down valence quark is heavier by about

(ms - md) (md/ms)^2 a(w) |Vds| =

= 312 x 0.25 x 0.253 x 0.22 Mev = 4.3 Mev,

(where
a(w) = 0.253 is the geometric part of the weak force strength
and
|Vds| = 0.22 is the magnitude of the K-M parameter
mixing first generation down and second generation strange)

so that **the Quantum color force constituent mass Qmd**
**of the down quark is**

**Qmd = 312.75 + 4.3 = 317.05 MeV.**

Similarly,
the up quark Quantum color force mass increase is about

(mc - mu) (mu/mc)^2 a(w) |V(uc)| =

= 1777 x 0.022 x 0.253 x 0.22 Mev = 2.2 Mev,

(where
|Vuc| = 0.22 is the magnitude of the K-M parameter
mixing first generation up and second generation charm)

so that **the Quantum color force constituent mass Qmu of the up quark is**

**Qmu = 312.75 + 2.2 = 314.95 MeV.**

**The Quantum color force Neutron-Proton mass difference is**

**mN - mP = Qmd - Qmu = 317.05 Mev - 314.95 Mev = 2.1 Mev.**

**Since the electromagnetic Neutron-Proton mass difference is**

**roughly mN - mP = -1 MeV**

**the total theoretical Neutron-Proton mass difference is**

mN - mP = 2.1 Mev - 1 Mev = 1.1 Mev,

an estimate that is fairly close
to the experimental value of 1.3 Mev.

Note that in the equation

(ms - md) (md/ms)^2 a(w) |Vds| = 4.3 Mev

Vds is a mixing of down and strange by a neutral Z0,
compared to the more conventional Vus mixing by charged W.
Although real neutral Z0 processes are suppressed
by the GIM mechanism,
which is a cancellation of virtual processes,
the process of the equation is strictly a virtual process.

```
Note also that the K-M mixing parameter |Vds| is linear.
Mixing (such as between a down quark and a strange quark)
is a two-step process,
that goes approximately as the square of |Vds|:
First
the down quark changes to a virtual strange quark,
producing one factor of |Vds|.
Then, second,
the virtual strange quark changes back to a down quark,
producing a second factor of |Vsd|, which is approximately
equal to |Vds|.

Only the first step (one factor of |Vds|) appears in the
Quantum mass formula used to determine the neutron mass.
If you measure the mass of a neutron,
that measurement includes a sum over a lot of histories
of the valence quarks inside the neutron.
In some of those histories, in my view,
you will "see" some of the two valence down quarks
in a virtual transition state that is at a time
after the first action, or change from down to strange,
and
before the second action, or change back.
Therefore, you should take into account
those histories in the sum in which you see a strange valence quark,
and you get the linear factor |Vds| in the above equation.


Note also that
if there were no second generation fermions,
or
if the second generation quarks had equal masses,
then
the proton would be heavier than the neutron
(due to the electromagnetic difference)
and
the hydrogen atom would decay into a neutron,
and
there would be no stable atoms in our world.
```

In this model, protons decay by virtual Black Holes over 10^64 years, according to by Hawking and his students who have studied the physical consequences of creation of virtual pairs of Planck-energy Black Holes.

---

## UCC - DCC Baryon Mass Difference:

According to a 14 June 20002 article by Kurt Riesselmann in Fermi News: "... The four [ first and second generation ] flavors - up, down, strange, charm - allow for twenty different ways of putting quarks together to form baryons ... Protons, for example, consist of two up quarks and one down quark (u-u-d), and neutrons have a u-d-d quark content. Some combinations exist in two different spin configurations, and the SELEX collaboration believes it has identified both spin levels of the u-c-c baryon. ... Physicists expect the mass difference between u-c-c and d-c-c baryons to be comparable to the difference in proton (u-u-d) and neutron (u-d-d) mass, since this particle pair is also related by the replacement of an up by a down quark. **The proton-neutron mass splitting**, however, **is sixty times smaller than the mass difference between the Xi_cc candidates** observed by the SELEX collaboration. ...

... Other questions, however, remain as well. The SELEX collaboration is puzzled by the high rate of doubly charmed baryons seen in their experiment. As a matter of fact, most scientists believed that the SELEX collaboration wouldn't see any of these particles. ...".

```
An up valence quark, constituent mass 313 Mev,
can swap places with a 2.09 Gev charm sea quark.
Therefore the Quantum color force constituent mass
of the down valence quark is heavier by about

  (mc - mu)    a(w)    |Vds|         =

=   1,777  x  0.253  x  0.22   Mev   =   98.9 Mev,

(where
a(w) = 0.253 is the geometric part of the weak force strength
and
|Vuc| = 0.22 is the magnitude of the K-M parameter
        mixing first generation up and second generation charm)

so that the Quantum color force constituent mass Qmu
of the up quark is
```

```
                       Qmu = 312.75 + 98.9 = 411.65 MeV.


A 313 Mev down valence quark can swap places
with a 625 Mev strange sea quark.
Therefore the Quantum color force constituent mass
of the down valence quark is heavier by about

  (ms - md)  a(w)    |Vds|         =

=   312  x  0.253  x  0.22   Mev   =   17.37 Mev,

(where
a(w) = 0.253 is the geometric part of the weak force strength
and
|Vds| = 0.22 is the magnitude of the K-M parameter
        mixing first generation down and second generation strange)

so that the Quantum color force constituent mass Qmd
of the down quark is
                       Qmd = 312.75 + 17.37 = 330.12 MeV.
```

Note that at the energy levels at which ucc and dcc live, the ambient sea of quark-antiquark pairs has at least enough energy to produce a charm quark, so that in the above equations there is no mass-ratio-squared suppression factor such as (mu/mc)^2 or (md/ms)^2, unlike the case of the calculation of the neutron-proton mass difference for which the ambient sea of quark-antiquark pairs has very little energy since the proton is almost stable and the neutron-proton mass difference is, according to experiment, only about 1.3 MeV.

Note also that these rough calculations ignore the electromagnetic force mass differentials, as they are only on the order of 1 MeV or so, which for ucc - dcc mass difference is small, unlike the case for the calculation of the neutron-proton mass difference.

```
The Quantum color force ucc - dcc mass difference is
```

mucc - mdcc = Qmu - Qmd = 411.65 MeV - 330.12 MeV = 81.53 MeV

Since the experimental value of the neutron-proton mass difference is about 1.3 MeV, the theoretically calculated

ucc - dcc mass difference is about 81.53 / 1.3 = 62.7 times the experimental value of the neutron-proton mass difference,

which is consistent with the SELEX 2002 experimental result that: "... The proton-neutron mass splitting ... is sixty times smaller than the mass difference between the Xi_cc candidates ...".

---

## Root Vector Geometry of Fermions, Spacetime, Gauge Bosons, and D4-D5-E6-E7-E8.

The 8 first-generation fermion particles can be represented as 8 vertices of a 24-cell

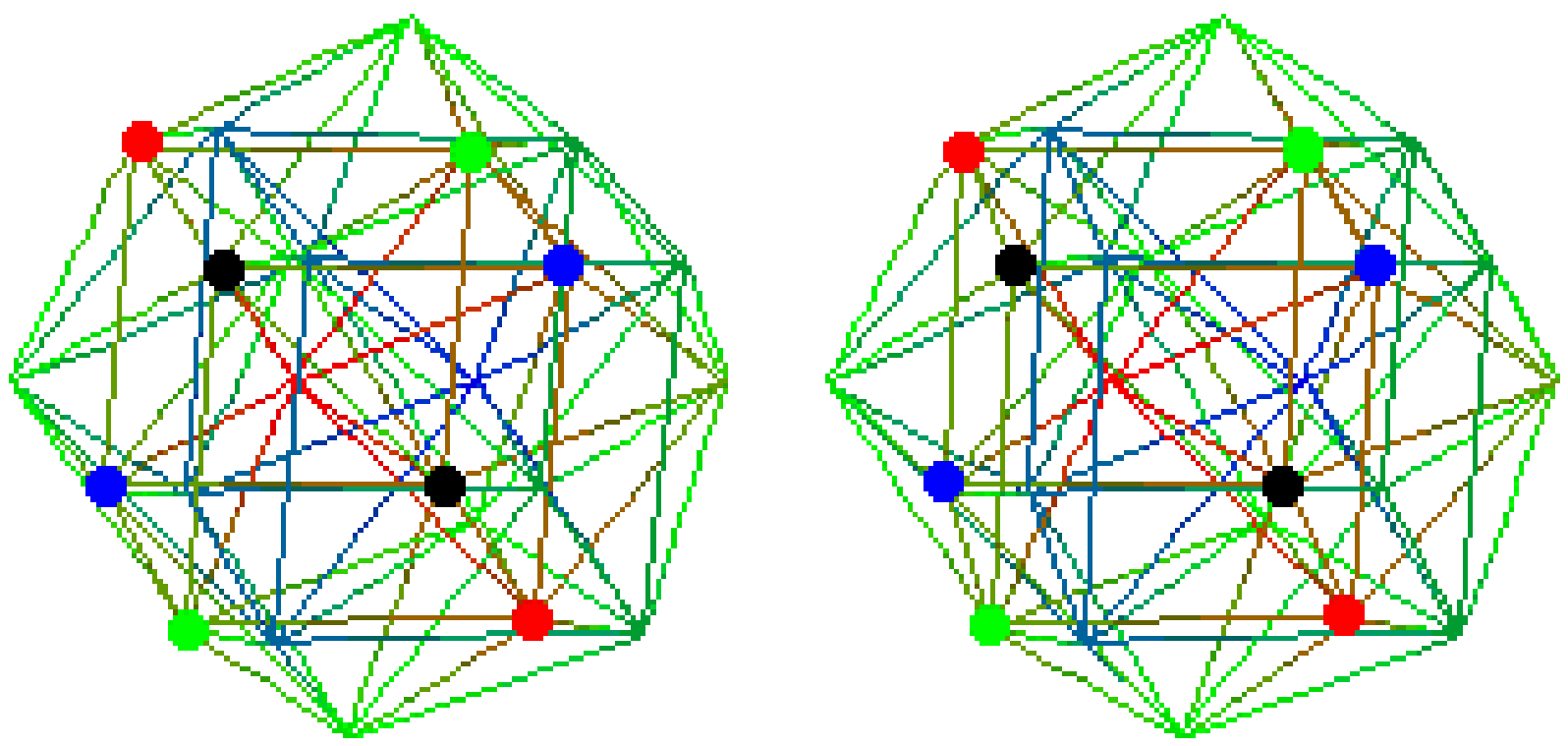

The 8 dimensions of unreduced spacetime (which reduces to 4-dimensional physical spacetime plus 4-dimensional internal Symmetry Space) can be represented as another 8 vertices of the 24-cell

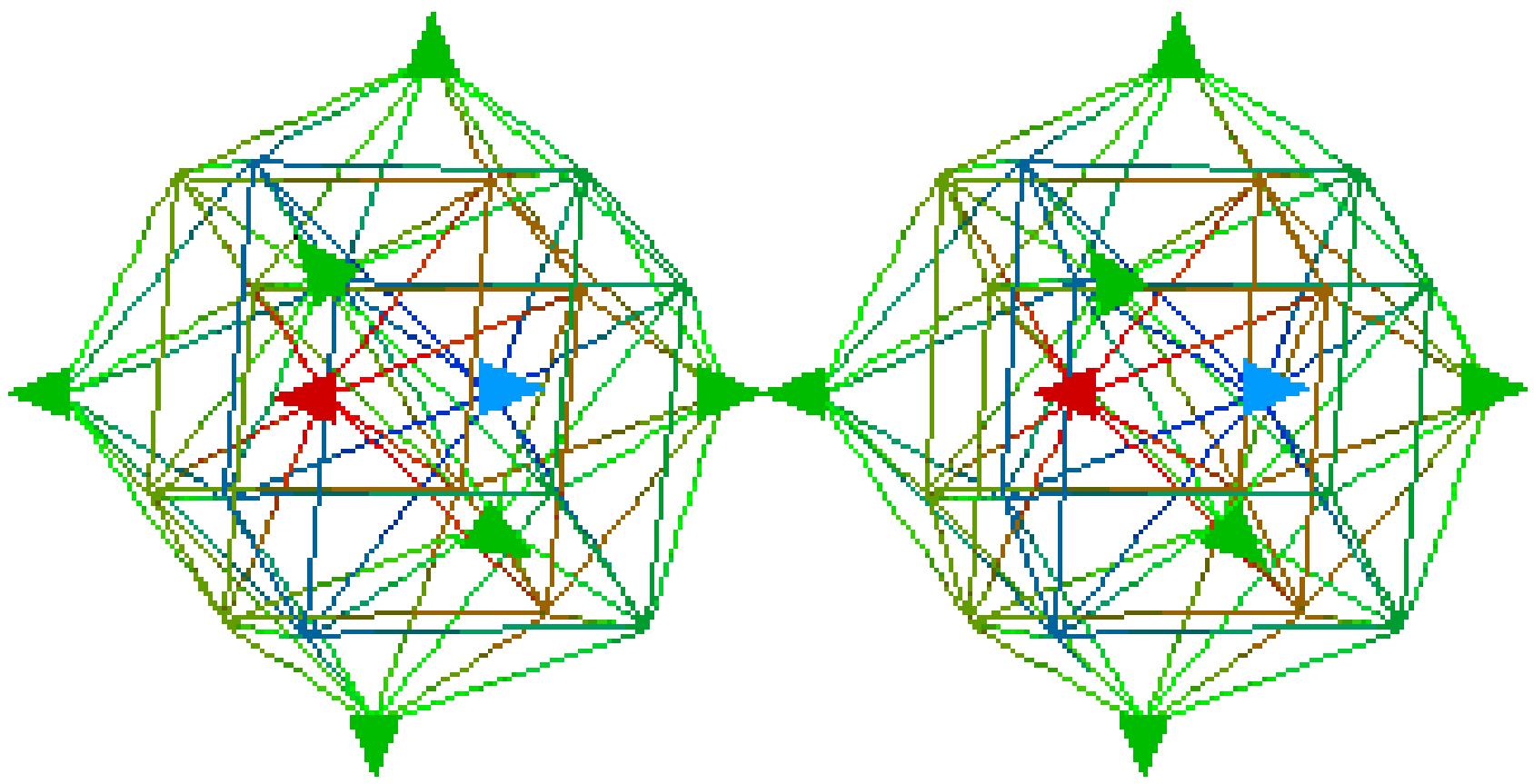

The third set of 8 vertices of the 24-cell then represents the 8 first-generation fermion particles, so that the entire 24-cell representation looks like

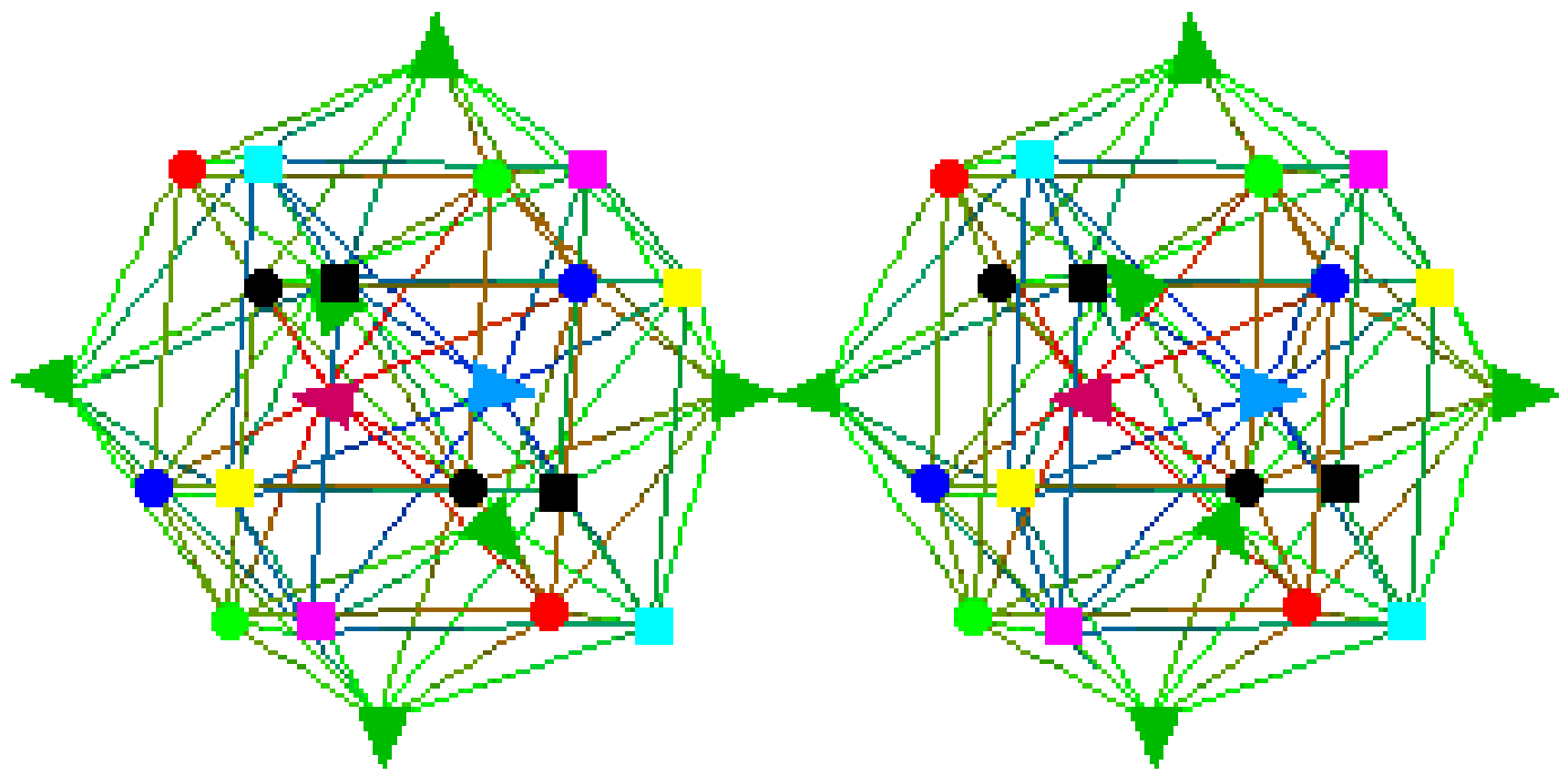

Note that the three sets of 8 vertices correspond to the two half-spinor and the vector representations of the D4 Lie Algebra are related by triality.

These relationships can also be viewed from the perspective of the Cl(1,7) Clifford Algebra structures

**1 8 28 56 70 56 28 8 1 = (8+8)x(8+8)**

As to the 28-dimensional adjoint representation, 24 of the 28 gauge boson D4 generators can be represented by the vertices of a dual 24-cell:

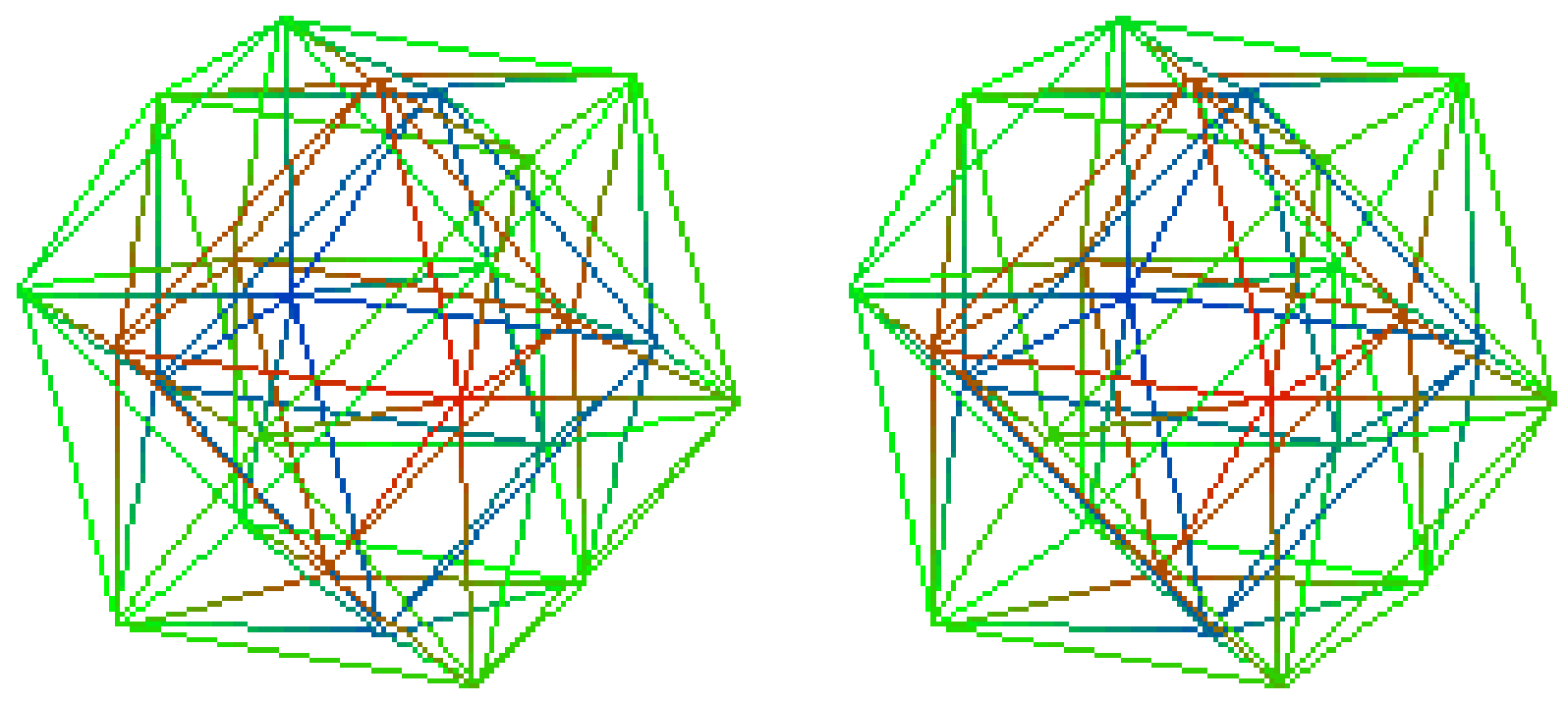

Note that the 24 + 24 = 48 vertices of the two dual 24-cells are 48 of the 72 root vector vertices of the E6 Lie algebra, and correspond to the 48 root vector vertices of the F4 subalgebra of E6:

> The 24 adjoint gauge boson vertices correspond to the 24 root vector vertices of the D4 subalgebra of E6;

When the 8 vector spacetime vertices are added, you get the 32 root vector vertices of the B4 subalgebra of E6;

When the 8+8 = 16 spinor vertices are added, you get the 48 root vector vertices of the F4 subalgebra of E6.

Here is how the 24 adjoint gauge boson vertices break down after dimensional reduction to form U(2,2) for gravity plus SU(3)xSU(2)xU(1) for the Standard Model:

12 of the 24 vertices correspond to the 12 vertices of the cuboctahedron that is the root vector polytope of the A3 = D3 Lie Algebra SU(2,2) = Spin(2,4) of the 4-dimensional Conformal Group. Then add the 4 D4 Cartan subalgebra generators to get the 12+4 = 16-dimensional Lie Algebra SU(2,2)xU(1) = Spin(2,4)xU(1) = U(2,2) that, by a generalization of the MacDowell-Mansouri mechanism, produces Gravity and the Higgs mechanism, and

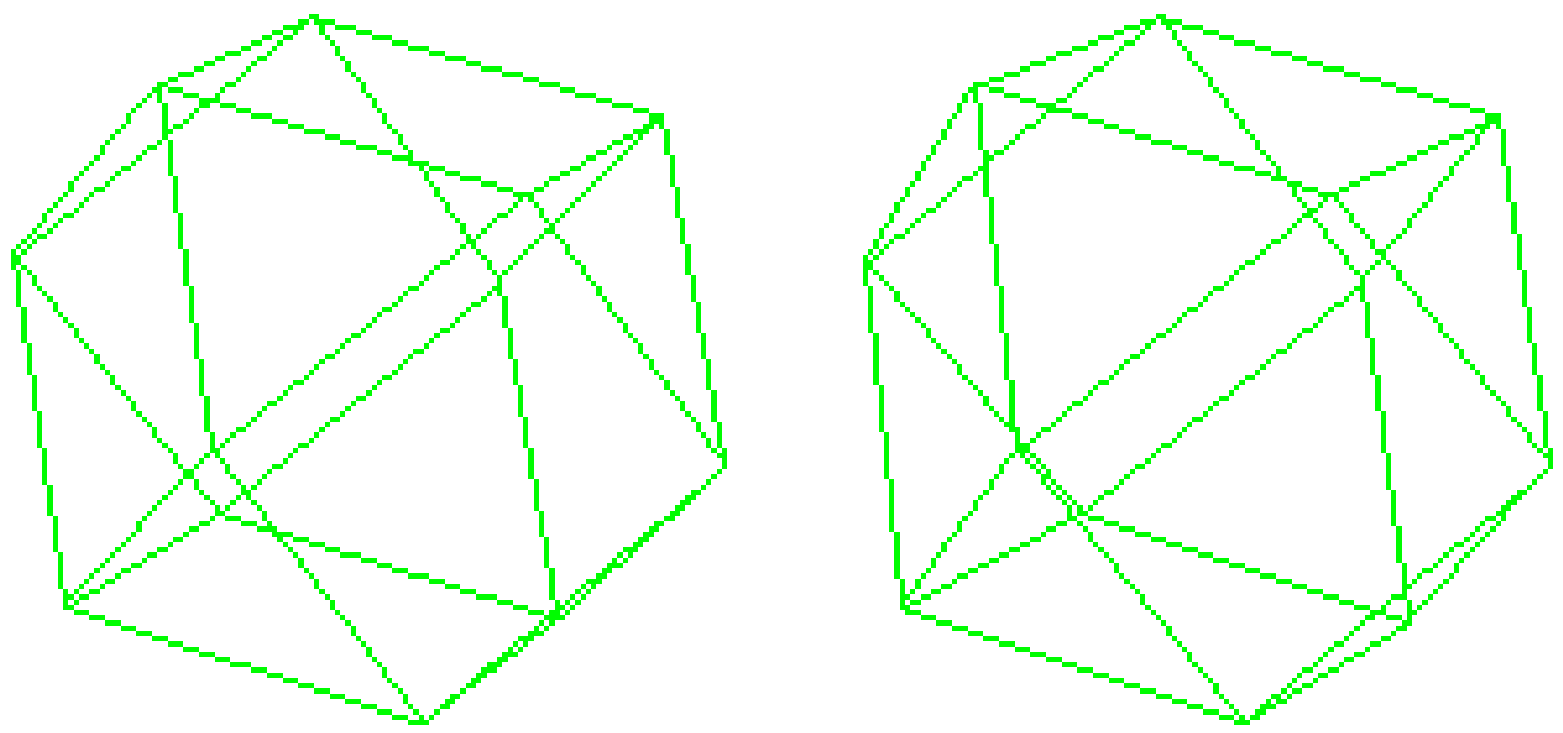

## 12 vertices of the 24 adjoint gauge boson vertices, plus the 4 D4 Cartan Subalgebra generators, represent the 16-dimensional U(2,2) for construction of Gravity plus Higgs.

That leaves 24-12 = 12 remaining vertices

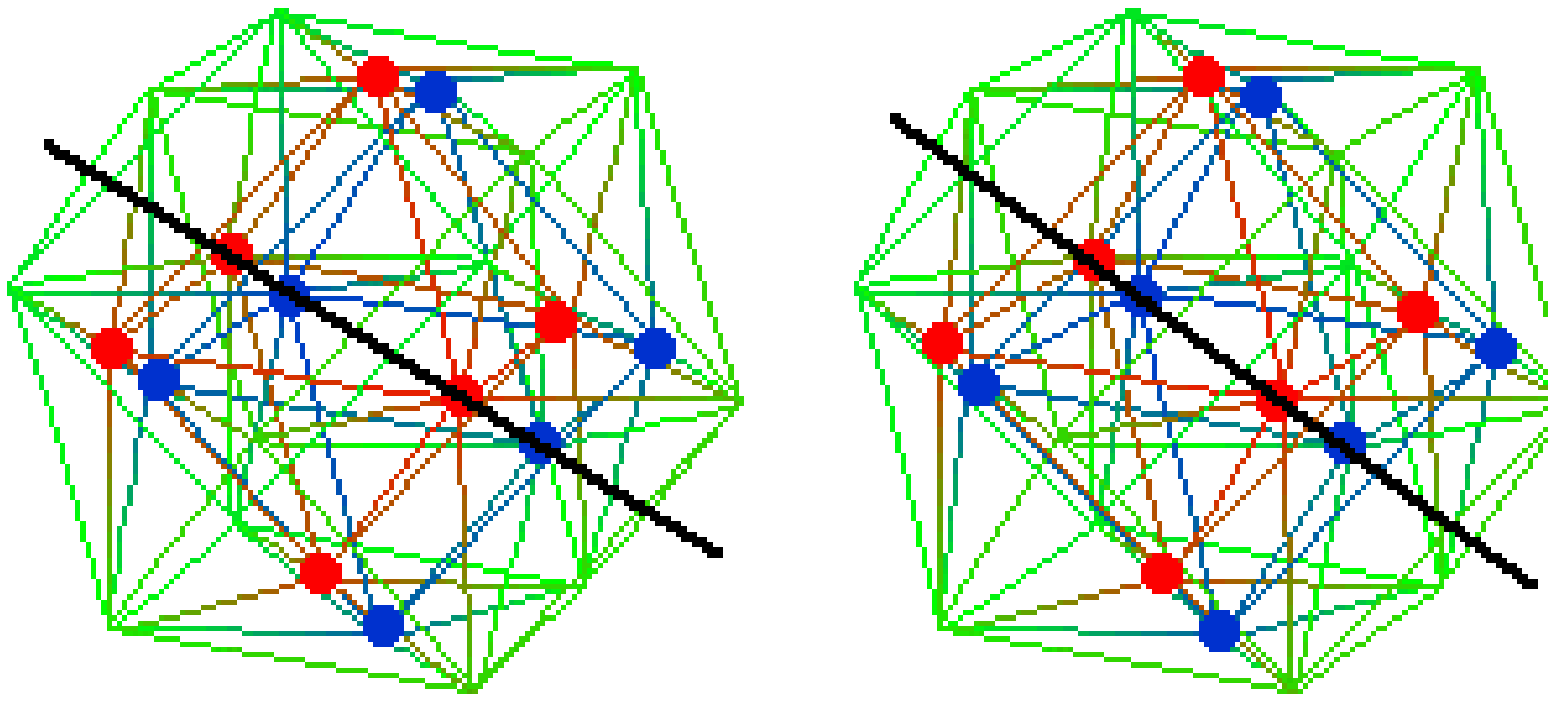

4 of which lie on a common line

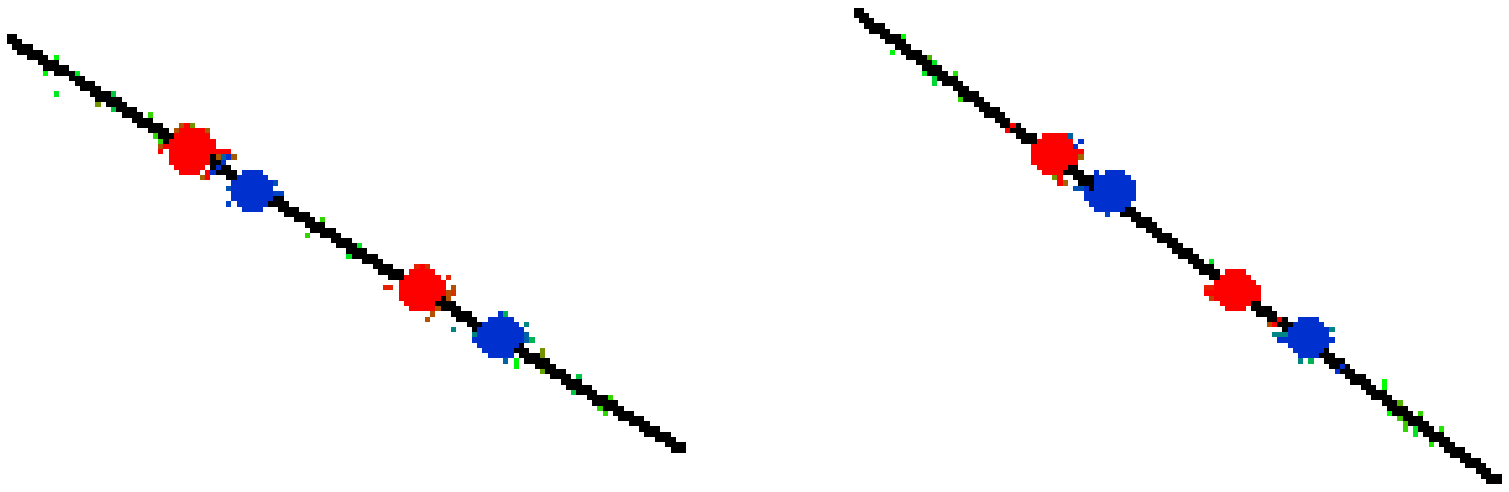

and represent the generators of the 4-dimensional Lie Algebra U(2) = SU(2)xU(1).

The remaining 8 vertices

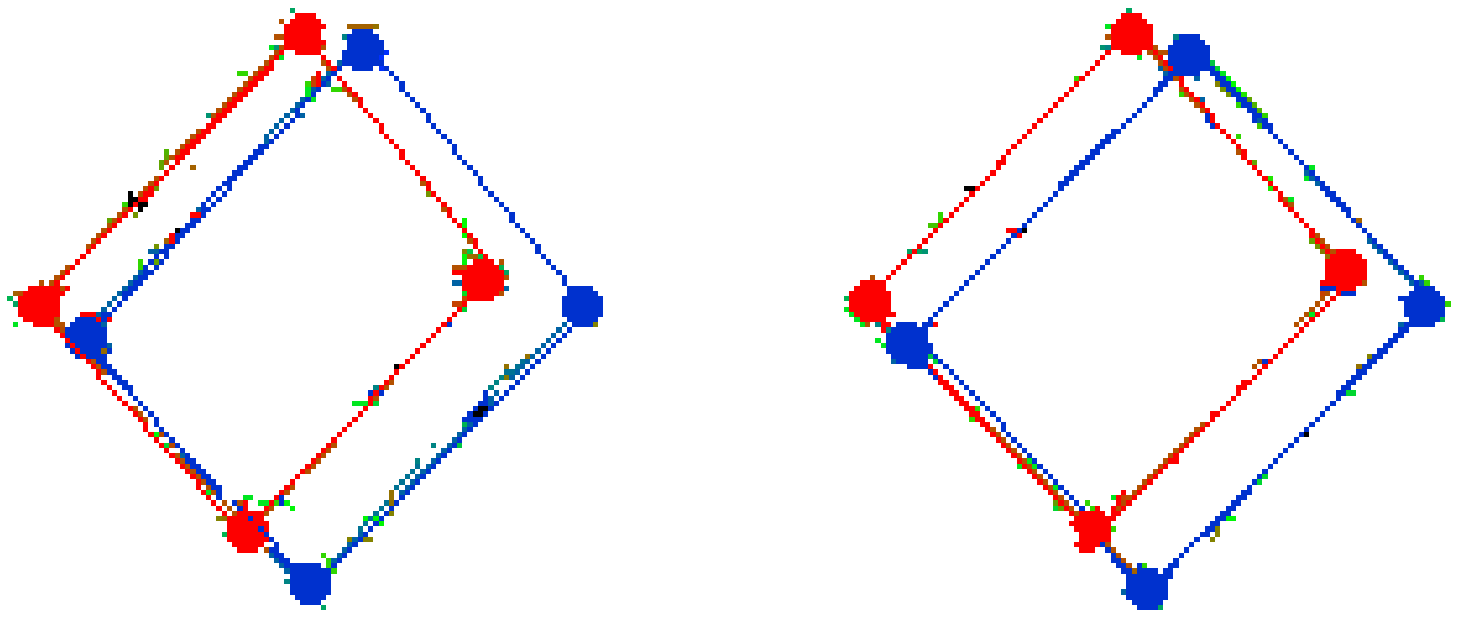

form a cube that can be labelled

```
    tb----xb
    |\    | \
   | zb----yb
    | |   |  |
  yr-|--zr  |
     \|     \|
```

```
xr----tr
```

Now look at the cube along its tb-tr diagonal axis, and project all 8 vertices onto a plane perpendicular to the tb-tr axis, giving the diagram

```
                 yb

         xb              zb

               tb  tr

         zr              xr

                 yr
```

with two central points surrounded by two interpenetrating triangles, which is the root vector diagram of SU(3), Therefore:

the 12 remaining vertices of the 24 adjoint gauge boson vertices represent the Standard Model Gauge Group SU(3xSU(2xU(1),

There is a nice geometric way to see the structures of D4-D5-E6-E7-E8 Lie Algebras:

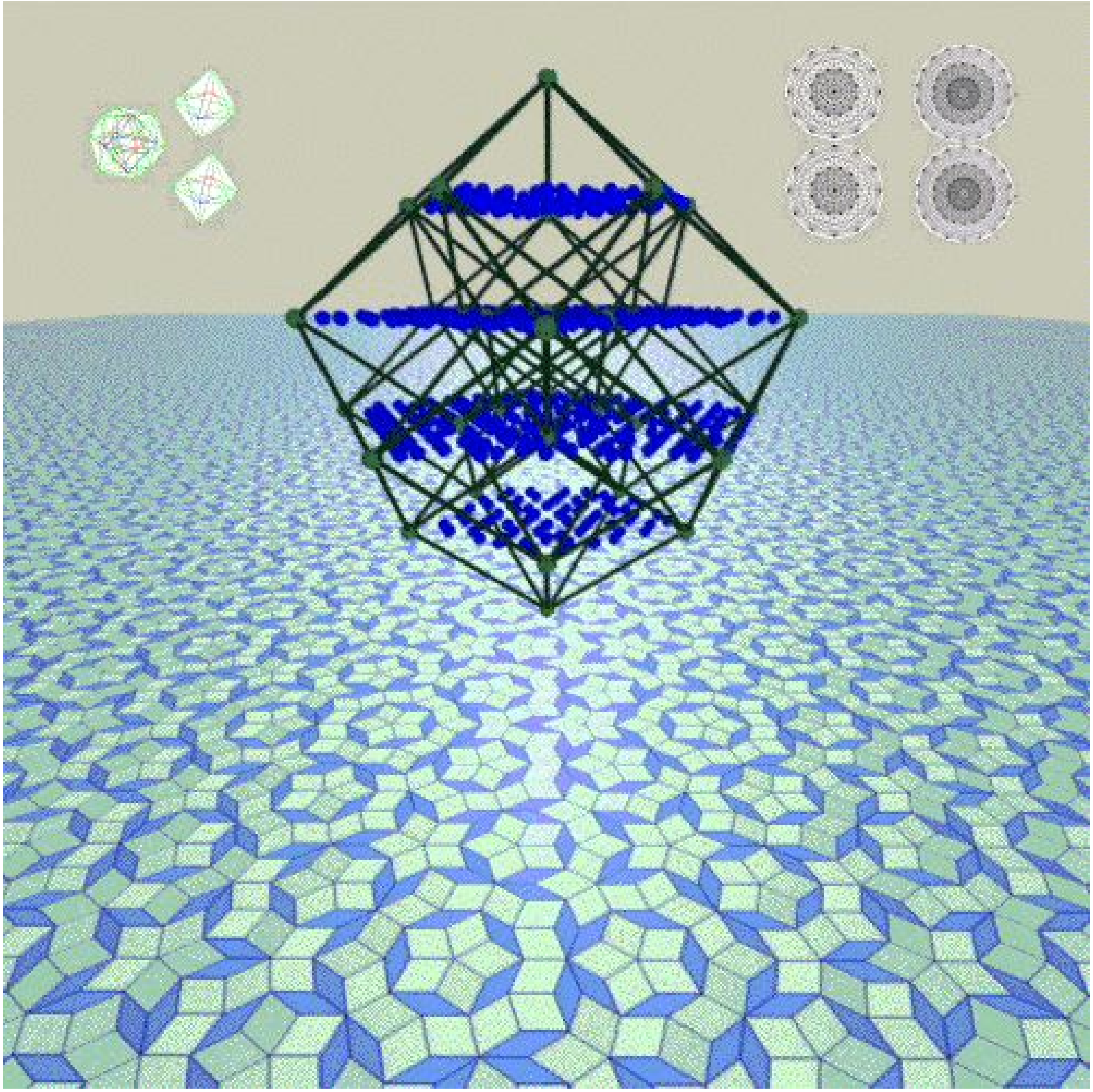

Floating above the Penrose-tiled plane in the above image (adapted from Quasitiler) are, going from left to right:

- 4-dimensional 24-cell, whose **24 vertices are the root vectors of the 24+4 = 28-dimensional D4 Lie algebra**;
- two 4-dimensional HyperOctahedra, lying (in a 5th dimension) above and below the 24-cell, whose **8+8 = 16 vertices add to the 24 D4 root vectors to make up the 40 root vectors of the 40+5 = 45-dimensional D5 Lie algebra**;
- 5-dimensional HyperCube, half of whose 32 vertices are lying (in a 6th dimension) above and half below the 40 D5 root vectors, whose **16+16 = 32 vertices add to the 40 D5 root vectors to make up the 72 root vectors of the 72+6 = 78-dimensional E6 Lie algebra**;
- two 27-dimensional 6-dimensional figures, lying (in a 7th dimension) above and below the the 72

E6 root vectors, whose 27+27 = 54 vertices add to the 72 E6 root vectors to make up the 126 root vectors of the 126+7 = 133-dimensional E7 Lie algebra; and

- two 56-dimensional 7-dimensional figures, lying (in an 8th dimension) above and below the the 126 E7 root vectors, and two polar points also lying above and below the 126 E7 root vectors, whose56+56+1+1 = 114 vertices add to the 126 E7 root vectors to make up the 240 root vectors of the 240+8 = 248-dimensional E8 Lie algebra.

E6 is an exceptional simple graded Lie algebra of the second kind:

E6 = g = g-2 + g-1 + g0 + g1 + g2

g0 = so(1,7) + R + iR

dim g-1 = 16

dim g-2 = 8

This gives real Shilov boundary geometry of S1xS7 for (1,7)-dimensional high-energy spacetime representation and for the first generation half-spinor fermion representations, which is the local structure needed for a local Lagrangian and calculation of ratios of particle masses and force strengths.

**Geometric Structure of NonLocal Quantum Theory is given by E6, E7, and E8:**

- **26-dim String Theory** with Jordan Algebra structure **J3(O)o** and Lie Algebra structure **E6 / F4** describes Generalized Bohm Quantum Theory with NonLinear Back-Reaction.
- **27-dim M-theory** with Jordan Algebra structures **J3(O) and J4(Q)o** and Lie Algebra structure **E7 / E6xU(1)** describes Timelike Branes in the MacroSpace of Many-Worlds.
- **28-dim F-theory** with Jordan Algebra structure **J4(Q)** and Lie Algebra structure **E8 / E7xSU(2)** describes Spacelike Branes in the MacroSpace of Many-Worlds.

---

# References:

The theoretical physics model described in this paper is not only based on the Lie Algebras D4, D5, E6, E7,

and E8, but also on the 256-dimensional Clifford Algebra Cl(1,7), whose 256 dimensions correspond to the 256 Odu of IFA, also known as Vodou. Therefore, the name that I prefer for this theoretical physics model is

## the D4-D5-E6-E7-E8 VoDou Physics Model.

Many details and references can be found on my home page on the web at URL

http://www.innerx.net/personal/tsmith/TShome.html

and on pages linked therefrom.

A few specific outside references are:

Particle Properties from the Particle Data Group at LBL

Quasitiler - The Penrose Tiling and 5-dim HyperCube in the image at the top of this page is from a web page of Alex Feingold. Unfortunately, the link to Quasitiler at the Geometry Center of the University of Minnesota is no longer good, because not only was the URL http://freeabel.geom.umn.edu/ changed to http://www.geom.umn.edu/ some time ago, but now the Geometry Center has gone away (NSF funds were cut off). Some of its pages live on in google cache, such as http://www.google.com/search?q=cache:s_WVB9Mv-4MC:www.geom.umn.edu/closed.html+&hl=en , but I don't know how to access all of them, and I feel that the loss of the Geometry Center is a loss to all of us who use the web.

---

Tony's Home

...